\documentclass[openany, oneside]{aa}  

\usepackage{float}
\usepackage{adjustbox}
\usepackage{csquotes}
\usepackage{graphicx}
\usepackage{rotating}
\usepackage{tabularx}
\usepackage{pdflscape}
\usepackage{rotating}
\usepackage{tablefootnote}
\usepackage{marvosym}
\usepackage{txfonts}
\usepackage{hyperref}
\usepackage{subcaption}
\hypersetup{
    colorlinks=true,
    linkcolor=blue,
    filecolor=magenta,
    citecolor=blue,
    urlcolor=blue,
    pdftitle={Overleaf Example},
    pdfpagemode=FullScreen,
    }
\begin{document}

   \title{Comprehensive study of five candidate $\delta$ Scuti-type pulsators\\ in detached eclipsing binaries}


\author{T. Pawar,\inst{1}\fnmsep\thanks{E-mail: pawartilak7@gmail.com}
\and
K. G. He{\l}miniak\inst{1}
\and
A. Moharana\inst{1}
\and
G. Pawar\inst{1}
\and
M. Pyatnytskyy\inst{2}
\and
H.N. Lala\inst{3}
\and
M. Konacki\inst{1}
}

\institute{Nicolaus Copernicus Astronomical Center, Polish Academy of Sciences, ul. Rabia\'{n}ska 8, 87-100 Toru\'{n}, Poland
\and 
Private Observatory \enquote{Osokorky}, PO Box 27, 02132 Kyiv, Ukraine
\and
Trivago N.V., Bennigsen-Platz 1, 40474 D\"usseldorf, Germany
        }

   \date{Received ** 2024; accepted ** 2024}

\abstract
{Pulsating stars in eclipsing binaries (EBs) provide an excellent opportunity to obtain precise, model-independent stellar parameters for studying these oscillations in detail. One of the most common classes of pulsators found in such EBs exhibits $\delta$ Scuti-type oscillations. Characterising these pulsators using the precise stellar parameters obtained using EB modelling can help us better understand such stars, and provide strong anchors for asteroseismic studies.}
   {We performed a comprehensive photometric and spectroscopic analysis of candidate pulsators in detached EBs, to add to the sample of such systems with accurately determined absolute parameters.}
   {We performed radial velocity and light curve modelling to estimate the absolute stellar parameters, and detailed spectroscopic modelling to obtain the global metallicity and temperatures. Frequency power spectra were obtained using residuals from binary modelling. Finally, we used isochrones to determine the age of the stars, and compared the estimated physical parameters to the theoretically obtained values.}
   {We present a detailed analysis of four candidate $\delta$ Scuti-type pulsators in EBs, and update the light curve analysis of the previously studied system TIC~308953703. The masses and radii of components are constrained to a high accuracy, which helps us constrain the age of the systems. We perform a Fourier analysis of the observed oscillations, and try to explain their origin. For TIC~81702112, we report tidal effects causing amplitude variation in the oscillation frequencies over the orbital phase.}

   \keywords{binaries: eclipsing -- stars: oscillations
                 -- variables: delta Scuti -- asteroseismology --
                stars: individual (HD~97329, V~Cir, HR~2214, V1109~Cen, HIP~7666)
               }

   \maketitle
%
\captionsetup[table]{font={rm,small}}
\section{Introduction}
Eclipsing binaries (EBs) are a sub-class of stellar binaries for which the orbital plane is along the observer's line of sight. They are crucial to the field of stellar astrophysics as they provide a myriad of information. Eclipses in their light curves (LCs) are affected by the stellar radii, the orbital inclination, the luminosities of the stars, and several other factors. Modelling these LCs can thus be seen as an inverse problem of finding the set of properties for these stars that reproduces the observed phenomena.
\par
However, LCs give us little to no information about the stellar masses. The mass of a star is arguably the most critical factor affecting its evolution. Double-lined EBs for which both components are visible in the observed spectra, provide us with the radial velocity (RV) semi-amplitudes, and hence the mass ratio and orbital elements, like the eccentricity and orbital separation between the stars. Further detailed analyses of the spectra can also anchor $T_{\mathrm{eff}}$, log\,$g$, metallicity, and individual chemical abundances.
\par
Detached EBs (DEBs) are such systems with well-separated stars, which have not undergone any interaction in the form of mass transfer. The stars in such systems are thus assumed to have evolved independently. Such systems are very important to paint an accurate picture of the age and evolution of stars by comparison with the theoretical models. They are also the gold standard when it comes to determining absolute stellar parameters as the accuracy of the derived parameters can be better than 1\% \citep{2019A&A...630A.106G, 2023MNRAS.520.1601K}.
\par
While EBs provide us with model-independent fundamental and atmospheric stellar parameters, our understanding of the structure and processes of the stellar interiors mostly remain model-dependent. However, the phenomenon of stellar pulsations has an intrinsic driving mechanism. Accurate modelling of observed pulsations thus provides us with a tool to probe the internal structure of stars. Pulsating stars in EBs come closest to offering both in a single system. A significant fraction of pulsating stars are being discovered in EBs as a consequence of the availability of high-precision photometry from space-based telescopes like {\it Kepler} \citep{2010Sci...327..977B} and the Transiting Exoplanet Survey Satellite \citep[TESS;][]{2015JATIS...1a4003R}.
\par
$\delta$ Scuti stars are A to F-type main-sequence (MS), pre-MS, or early post-MS stars, typically with masses ranging from 1.5 to 2.5 $M_{\mathrm{sun}}$, although in certain cases this lower limit can extend down to 1.3 $M_{\mathrm{sun}}$ \citep{2011AJ....142..110M}. They generally exhibit pressure mode oscillations with periods between 18\,min and 8\,h and amplitudes below $0^m.1$ in the V-band \citep{2010aste.book.....A}. Some of these systems have been studied before to understand the effect of binarity on stellar pulsations. Some exhibit tidally induced and tidally tilted pulsations and help us gauge the effect of tidal interaction \citep{2020MNRAS.498.5730F}. However, the list of such systems with comprehensive analysis remains short.
\par
In this study, we present a detailed analysis of five EBs with candidates for $\delta$ Scuti-type pulsators as one of the components. We explain the process of sample selection and obtaining the data. Analyses for this study can be broadly divided into the following sections: photometric analysis, RV and spectroscopic analysis, pulsation analysis, and evolutionary modelling. We then discuss the results and present our conclusions and summary.
\section{Target selection and observations}
The targets for this project were selected in various ways. First, by using data obtained from the catalogue of detached EBs generated by the CR\'EME survey \citep{2013MNRAS.433.2357R, 2014A&A...567A..64H, 2015MNRAS.448.1945H, 2019A&A...622A.114H, 2021MNRAS.508.5687H, 2022pas..conf..163H}. The survey meticulously monitored a vast sample of more than 300 EBs and acquired multi-epoch spectra to enhance our understanding of these systems. Using a pre-cut on the projected minimum masses, $M\sin^3(i)$, acquired from RV modelling, systems were selected in the mass range corresponding to that of $\delta$ Scuti-type pulsators. The LCs of these EBs were then visually inspected to look for pulsations. Three such systems are presented here: TICs~81702112 (HD~97329), 189784898 (V~Cir), and 165459595 (V1109~Cen). 

A similar approach was applied to targets from the catalogue DEBCat \citep{2015ASPC..496..164S}, with the focus on targets that only had ground-based photometric data published. In this work, we analyse one such system -- TIC~308953703 (HIP~7666). Finally, one of us (MP) recently reported eclipses and pulsations discovered in the LC of TIC~386622782, a previously known spectroscopic double-lined (SB2) binary HR~2214. After this announcement, the target was included in the CR\'EME sample, and scheduled for additional spectroscopic observations. 

\subsection{Photometry}
\label{sec:Photometry}
\begin{figure*}[htbp]
    \centering
    \includegraphics[width=0.95\textwidth]{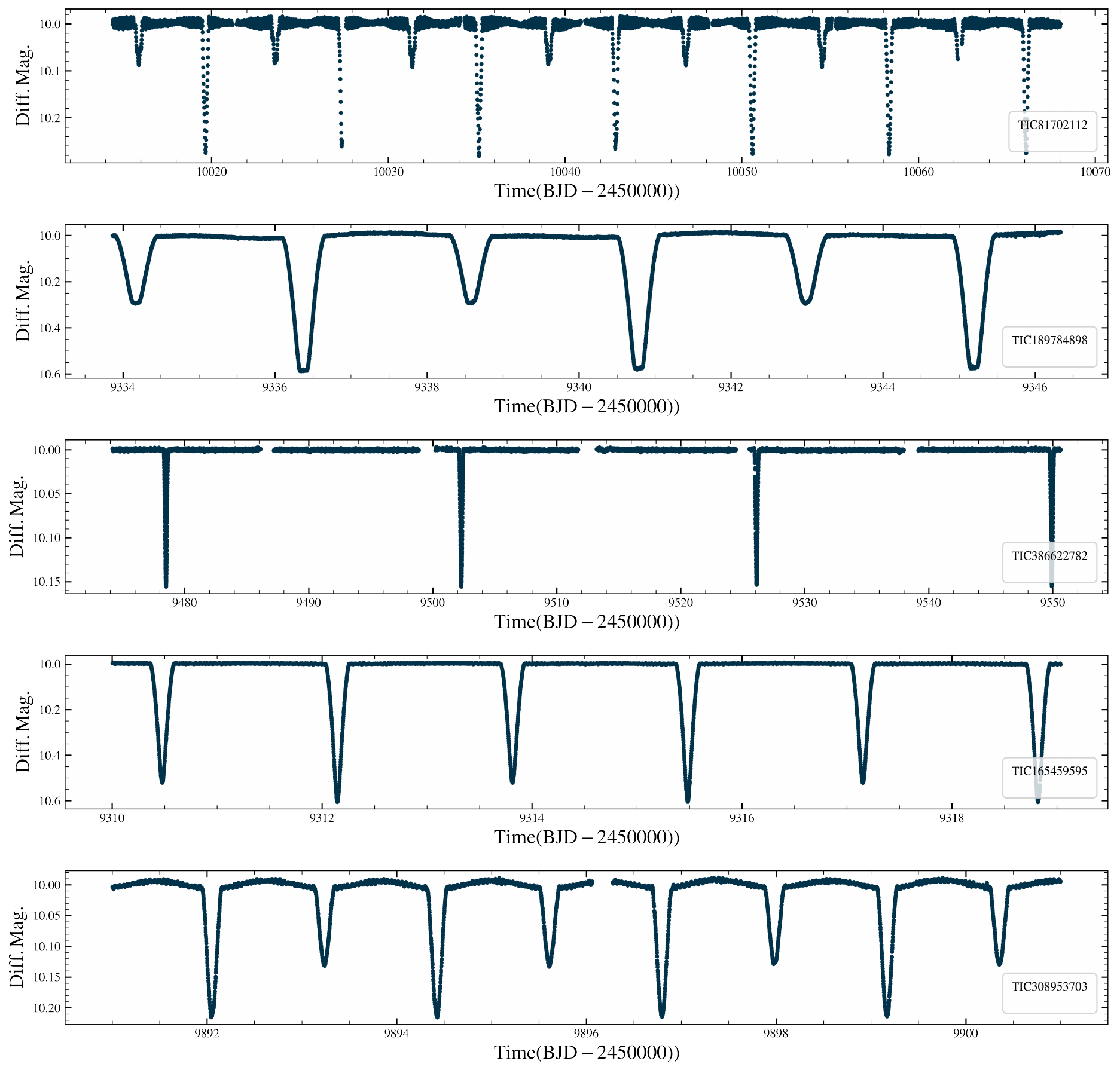}
    \caption{Short (2-min) cadence LCs of the targets, each for a single TESS sector.}
    \label{fig:LCs}
\end{figure*}

High-precision TESS photometry, in the form of 2-min cadence LCs, is available for all the targets in the sample.\footnote{Through Guest Investigator proposals: G011083, G03028, G05078, G06112 (PI: K.~He{\l}miniak), G011060 (PI: E.~Paunzen), G04123 (PI:~V.~Antoci), and G05003 (PI: A.~Pr\v{s}a), as well as the core science programme} These LCs are available through the Mikulski Archive for Space Telescopes (MAST) and were retrieved using the \textsc{lightkurve} package \citep{2018ascl.soft12013L}. This short-cadence photometry ensures a good resolution of the pulsation frequencies, especially beneficial for short-period pulsators like $\delta$ Scuti-type variables, and ensures a reliable model to estimate stellar masses and radii. We chose the simple aperture photometry (SAP) fluxes for all the targets to avoid distortions in the LCs introduced through Pre-search Data Conditioning (PDC)-SAP fluxes. The data was converted to differential magnitude and shifted to a baseline of 10 to facilitate the input file requirements of the LC modelling code. The choice of this baseline was arbitrary and does not affect the physical parameters in the solution. The LCs of all five targets used in this study are shown in Fig.~\ref{fig:LCs}

\subsection{High resolution spectra}
\label{sec:Spectroscopy}
Most of the spectra used for analysis were obtained using the CHIRON spectrograph \citep{2013PASP..125.1336T}, attached to the 1.5\,m SMARTS telescope installed at Cerro Tololo Inter-American Observatory (CTIO). Aiming for higher efficiency, the spectrograph was used in the fiber mode, providing a resolution of $\sim$28000. Extracted and wavelength-calibrated spectra were obtained with the pipeline developed at Yale University \citep{2013PASP..125.1336T} and provided to the user. However, barycentric velocity corrections were done in-house using the {\it bcvcor} procedure within IRAF \citep{1986SPIE..627..733T}. Forty echelle orders were combined and continuum corrected to be used for RV and spectral analysis. 

CHIRON was used to observe TICs 81702112, 189784898, 165459595, and 386622782. For the latter, our new data were complemented with the archival measurements from \citet{1985ApJS...59..229A}. We would like to point out that our observing programme on CHIRON was stopped abruptly when the Association of Universities for Research in Astronomy announced the shutdown of the SMARTS telescopes in September 2023. We would have gathered and presented more data, especially for TIC~16545959, if not for this unfortunate event.  

For TIC~189784898, we also obtained six spectra taken with the FEROS spectrograph at the MPG-2.2\,m telescope \citep{1999Msngr..95....8K}, providing a higher resolution of $48000$. These data were taken in 2004 during an engineering run, and are available from the ESO Archive. No new spectroscopic data was obtained for the target TIC$~$308953703. The RV measurements obtained by \citet{2021MNRAS.508..529F} were adopted while carrying out the RV+LC analysis. A short summary of the spectroscopic observations is provided in Table~\ref{tab:spectra log}.

\begin{table*}[htbp]
\small
    \centering
    \caption{Ground-based spectroscopic observations log.}
    \label{tab:spectra log}
    \begin{tabular}{cccccccc}
       TIC ID & Other ID & RA & Dec & $V_{\mathrm{mag}}$& GDR3$^*$ dist. & Instrument & No. \\
        & & (deg., ep=J2000) & (deg., ep=J2000) & & (pc) & &  \\
        \hline
         81702112 & HD 97329 & 167.892277 & $-$49.936389 & 8.31 &295.94& CHIRON & 10 \\   
        189784898 & V Cir & 221.228092 & $-$57.032231 & 10.76&548.67& CHIRON+FEROS & 8+6 \\
        386622782 & HR 2214 & 93.619095 & $+$17.906349 & 8.53 & - & CHIRON+\citet{1985ApJS...59..229A} & 9+16\\
        165459595 & V1109 Cen & 180.192040 & $-$40.354501 &9.60 & 310.16& CHIRON & 6 \\
        308953703 & HIP~7666 & 24.673619 & $+$52.518800 & 9.64& 313.43& \citet{2021MNRAS.508..529F} & 10  \\
        \hline
    \end{tabular}
\\ $^*$ \textit{Gaia} Data Release~3 (GDR3; \citep{2023A&A...674A...1G})
\end{table*}

\section{Spectroscopy}
\subsection{Radial velocities}
To calculate the RV values, we used the two-dimensional cross-correlation technique implemented in the \textsc{todcor} program \citep{1994ApJ...420..806Z}. Synthetic spectra used as templates were calculated using the ATLAS9 model atmosphere \citep{1979ApJS...40....1K}. The values of $T_\mathrm{eff}$, metallicity, and $v\sin(i)$ were taken as 6800\,K, 0.0, and 30\,km\,s$^{-1}$, respectively. Errors were calculated using a bootstrap approach \citep{2012MNRAS.425.1245H}.

Orbital solutions for the extracted RVs were calculated using {\small \textsc{v2fit}} \citep{2010ApJ...719.1293K}. The routine estimates the orbital parameters of a double-Keplerian orbit fitted to the data, and is minimised using a Levenberg-Marquardt scheme. In addition to this, simultaneous LC-RV modelling was performed for four of the five targets. The exception was TIC~81702112, due to high root mean square ($rms$) values of the secondary RV fluctuations leading to erroneous outcomes.

\begin{figure*}[h!]
\centering
\begin{subfigure}{0.49\textwidth}
  \centering
  \includegraphics[width=\linewidth]{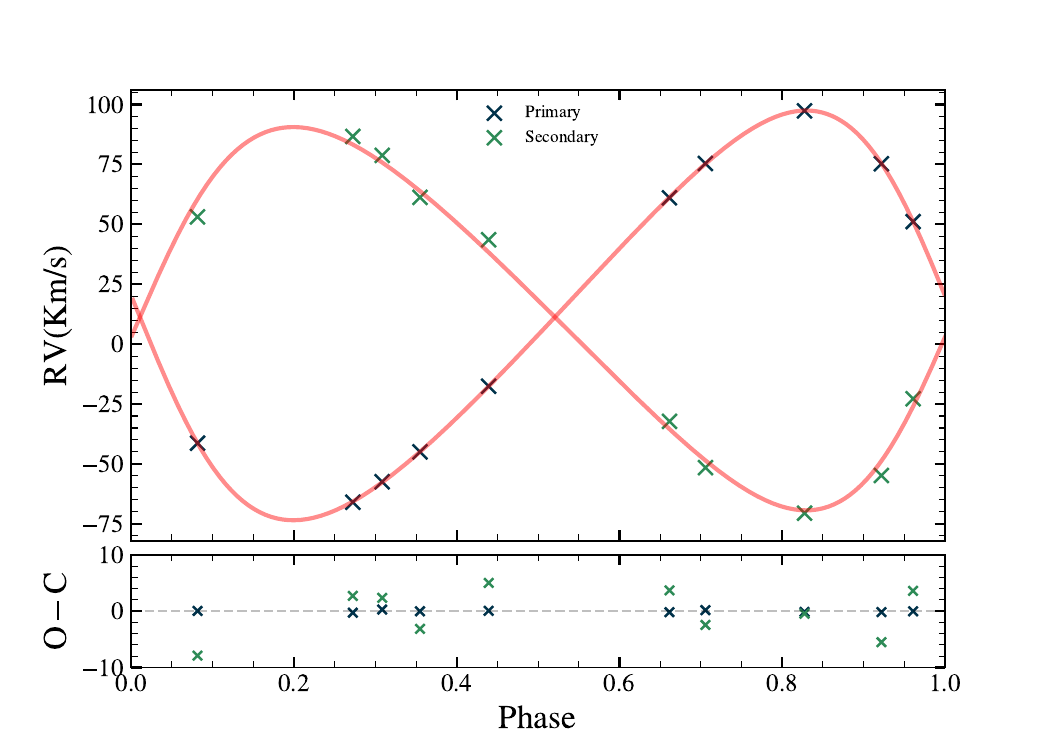}
  \caption{TIC~81702112}
\end{subfigure}
\hfill
\begin{subfigure}{0.49\textwidth}
  \centering
  \includegraphics[width=\linewidth]{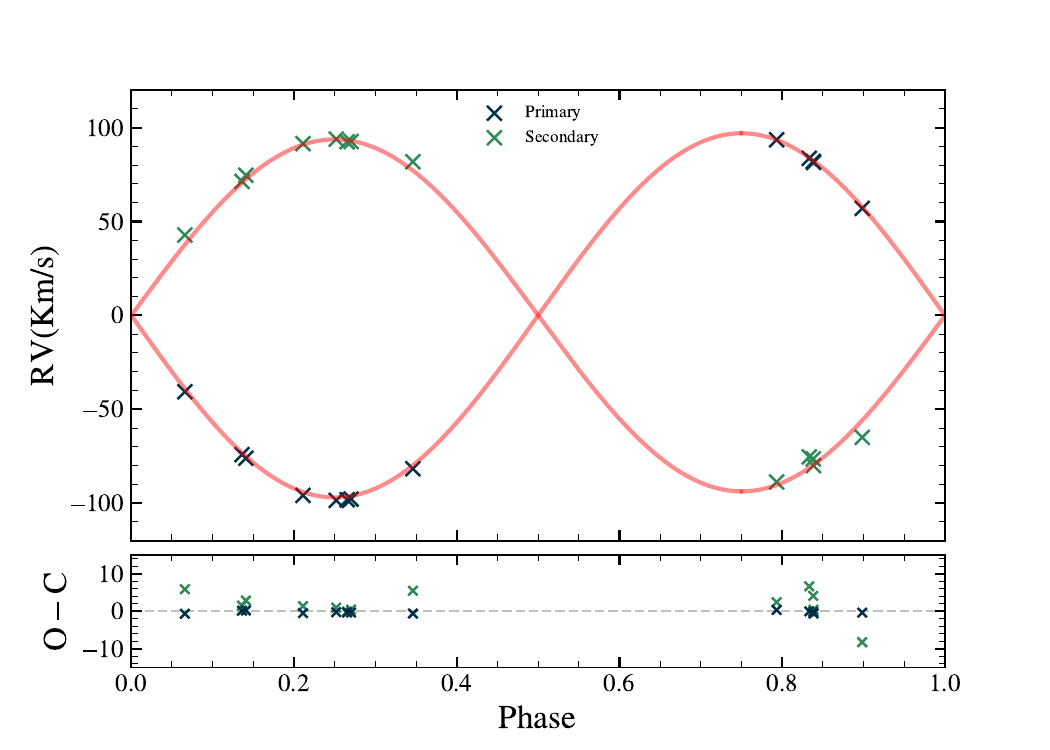}
  \caption{TIC~189784898}
\end{subfigure}
\centering
\begin{subfigure}{0.49\textwidth}
  \centering
  \includegraphics[width=\linewidth]{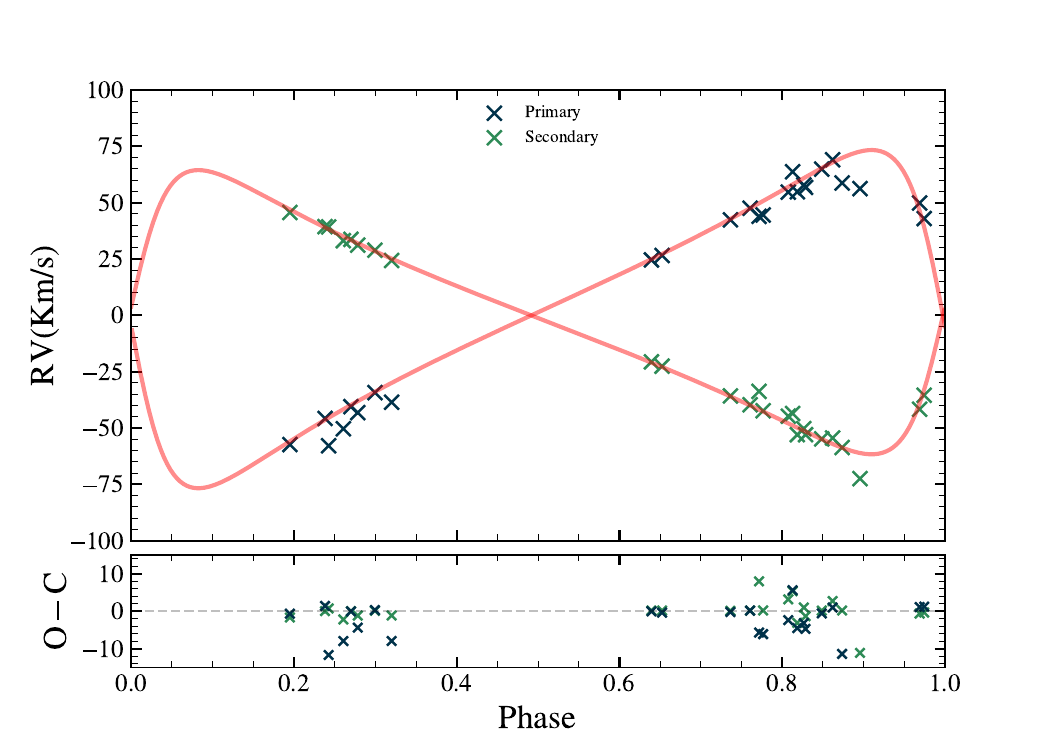}
  \caption{TIC~386622782}
\end{subfigure}
\hfill
\begin{subfigure}{0.49\textwidth}
  \centering
  \includegraphics[width=\linewidth]{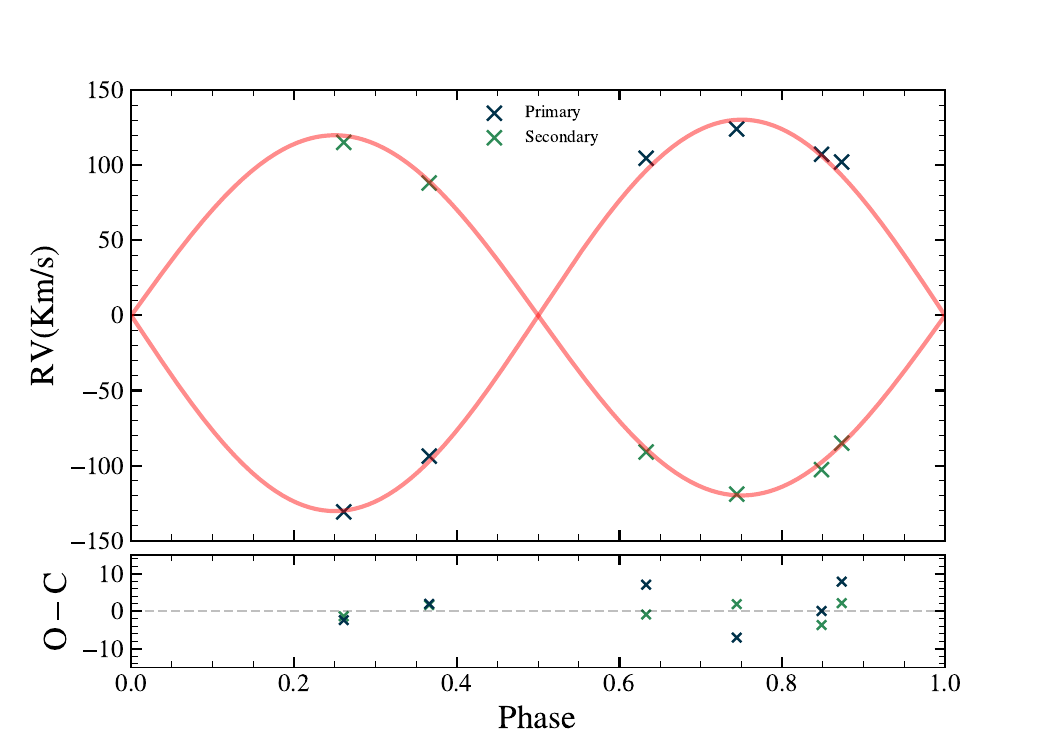}
  \caption{TIC~165459595}
\end{subfigure}
\hfill
\begin{subfigure}{0.49\textwidth}
  \centering
  \includegraphics[width=\linewidth]{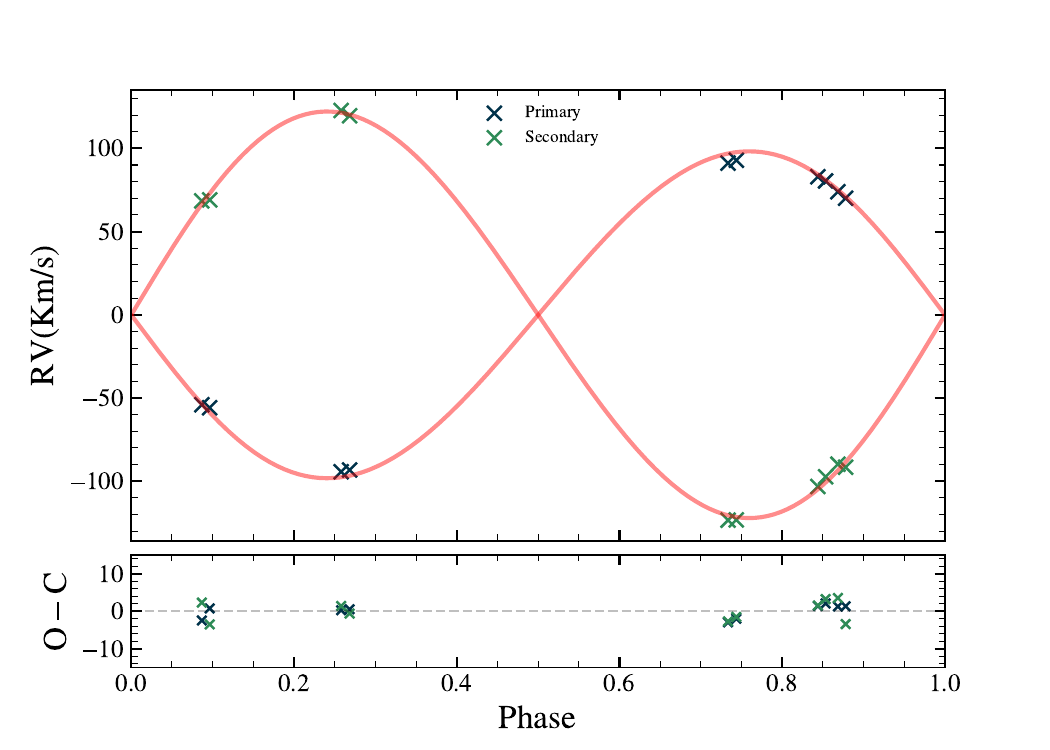}
  \caption{TIC$~$308953703}
\end{subfigure}

\caption{Orbital fits obtained for the RV curves for the targets. Here, primary refers to the more massive star and secondary is the less massive companion.}
\label{fig:RVFits}
\end{figure*}

We fit for the orbital period, $P\mathrm{_{orb}}$, zero-phase, $T\mathrm{_p}$, systemic velocity, $\gamma$, velocity semi-amplitudes, $K\mathrm{_{1,2}}$, eccentricity, $e$ and  longitude at periastron passage, $\omega$. The orbital fits are displayed in Fig.~\ref{fig:RVFits}.

\subsection{Broadening functions}
In the context of LC and spectral analyses, similar to any multi-parameter problem, it is advisable to obtain and fix as many parameters from various methods as possible to mitigate degeneracy. In the context of spectral disentangling and fitting, we resorted to broadening functions (BFs) to mainly obtain two necessary parameters: the rotational velocities of the stars and their light contributions.

The BF depicts the spectral profiles in velocity space, containing the signatures of RV shifts across various lines, along with intrinsic stellar phenomena such as rotational broadening, spots, and pulsations \citep{1999TJPh...23..271R}. The implementation of this method is described in detail by \citet{2023MNRAS.521.1908M}. The fit was obtained for all the spectra for which the peaks of both components were well separated, avoiding in- or near-eclipse epochs. The final parameter values were calculated by taking the average of the ones from individual fits. The line profiles are extremely noisy in the case of TIC~165459595, preventing us from performing BF analysis for this target. The obtained parameters are stated in Table~\ref{tab:BF output}.

\begin{table}[htbp]
   \small
    \centering
    \caption{BF output}
    \label{tab:BF output}
    \begin{tabular}{cccc}
       System & $v\sin(i)$ (Pri.) & $v\sin(i)$ (Sec.) & $L_\mathrm{sec}$/$L_\mathrm{pri+sec}$ \\
        \hline
        TIC 81702112 & 23.2$\pm$0.7 & 27.4$\pm$3.2 & 0.48 \\
        TIC 189784898 & 31.5$\pm$0.7 & 52.2$\pm$1.4 & 0.56 \\
        TIC 386622782 & 18.6$\pm$1.3 & 19.1$\pm$0.2  & 0.7 \\
        \hline
    \end{tabular}
\end{table}

\subsection{Spectral disentangling}
A lot can be inferred about a star by studying its spectra. Estimates of temperature and surface gravity obtained from spectral analysis can provide useful checks on the parameters, such as flux ratios and relative radii, obtained from the LC modelling. They also provide an estimate of the global metallicity, which is an important parameter when constraining age and evolution. However, spectra obtained for binaries need to be disentangled into separate components before we can proceed with the modelling.
\par
One way to do this involves obtaining spectra at different epochs in the orbit. Due to the different Doppler shifts, the spectrum corresponding to each star will have a different shift in the observed composite spectrum. By using a number of such spectra and the associated RVs, it is possible to reconstruct an individual component spectrum independent of any template.
\par
\begin{figure}[htbp]
         \centering
         \includegraphics[width=\columnwidth]{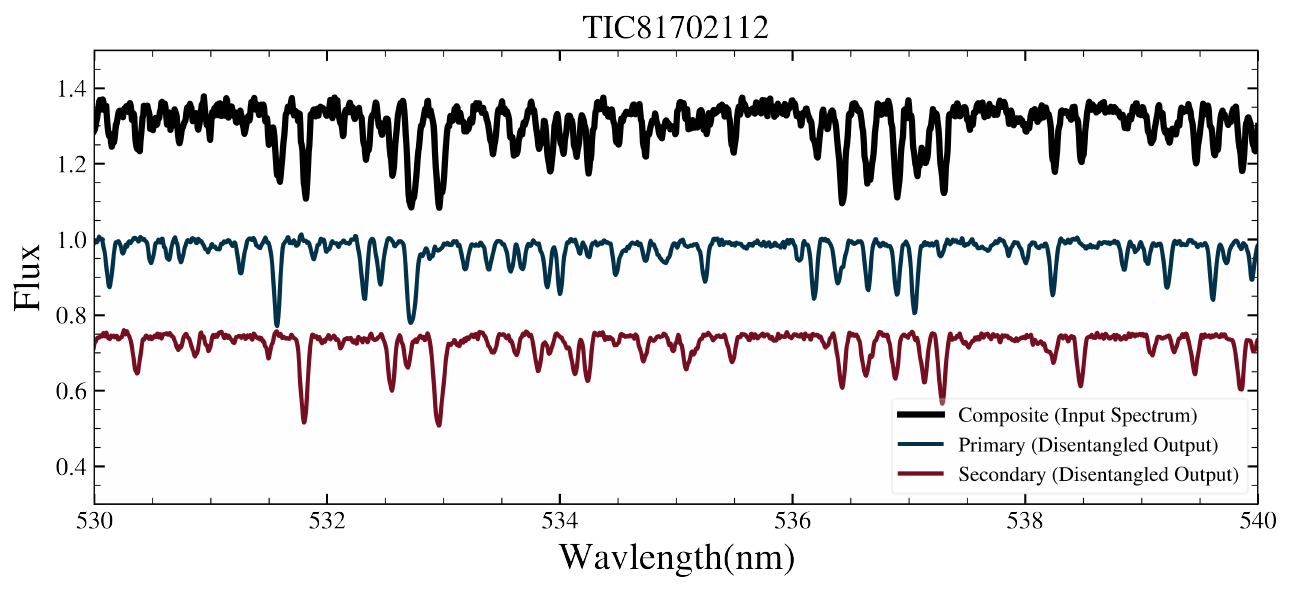}
         \caption{Input and output spectra utilised and obtained during the spectral disentangling for the target TIC~81702112. \textbf{Black}: One of the observed spectra that was used as an input for the disentangling algorithm. This is a composite of contributions from both the stars. \textbf{Blue}: Disentangled spectrum of the primary component. \textbf{Red}: Disentangled spectrum of the secondary component.
         }
         \label{fig:A111134_spectral_disentangling}
\end{figure}

One implementation of this technique to disentangle stellar spectra is shift-and-add \citep{2020A&A...639L...6S, 2020AN....341..628Q, 2006A&A...448..283G}. It is an iterative process whereby the component spectra are shifted according to the corresponding RVs and added at each iteration. This process uses a set of N spectra, each at a different epoch in the orbit. The algorithm begins by correcting the spectra of component A for the RVs and adds them up. This serves as a first approximation of the component A spectra. This co-added spectra is then subtracted from all the input spectra to obtain the first approximation of component B. This process is repeated for a large number of iterations in order to refine and improve the accuracy of the separation process. This process also improves the signal-to-noise (S/N) by the square root of N. Figure \ref{fig:A111134_spectral_disentangling} illustrates the spectral disentangling process for TIC~81702112, showing one of the input spectra alongside the resulting component spectra obtained after applying the shift-and-add algorithm.

To implement this algorithm, we used the code \textsc{disentangling\_shift\_and\_add} \citep{2020A&A...639L...6S, 2022A&A...665A.148S}. We executed the procedure for a wavelength range of 500-580 nm to avoid broad-wing features while ensuring that sufficient narrow lines with decent S/N were present. We used the RV semi-amplitudes obtained using \textsc{v2fit}. The flux ratio calculated using BFs was provided as an input, which was used by the routine to scale the disentangled components. For TIC~165459595, we used the flux ratio obtained from the \textsc{jktebop} solution. In case of TIC~189784898, we only used the six FEROS (R=48000) spectra obtained for this target. Due to the high S/N of the input spectra and higher spectral resolution, the individual decomposed spectra were of better quality than for the other cases. The procedure was run for 30,000 iterations in each case. 

The initial detrended spectrum has trends in the continuum as a result of the imperfections in the continuum normalisation of composite spectra, line profile variations etc. \citep{spuriouspatt}. These trends or biases are additive and can be removed by fitting the continuum and then subtracting it from the disentangled spectra. This gave us the final spectra that we used for spectral analysis.

\subsection{{Spectral analysis}}

We used the {\small \textsc{iSpec}} framework \citep{2014A&A...569A.111B,2019MNRAS.486.2075B} for the purpose of atmospheric parameter determination using the disentangled spectra. Further continuum normalisation within \textsc{iSpec} for example, using third-order splines was avoided to ensure that line depths remained least affected. Shifts in the RVs resulting from the disentangling process were removed before analysis. 

Within the synthetic spectral-fitting technique, \textsc{iSpec} provides users with the ability to generate synthetic spectra on the go using several choices of radiative transfer codes. We used the code \textsc{spectrum} \citep{1994AJ....107..742G} to generate synthetic spectra in the wavelength region of our interest, owing to its speed and accuracy. We used the line lists from the \textit{Gaia}-ESO Survey \citep[GES;][]{2015PhyS...90e4010H}, covering the wavelength range of 420-920 nm. \textsc{iSpec} uses $\chi^2$ minimisation to choose the best-fitting spectrum from the ones synthesised.

\begin{figure*}[htbp]
  \centering
  \begin{subfigure}[b]{0.925\linewidth}
    \includegraphics[width=\linewidth]{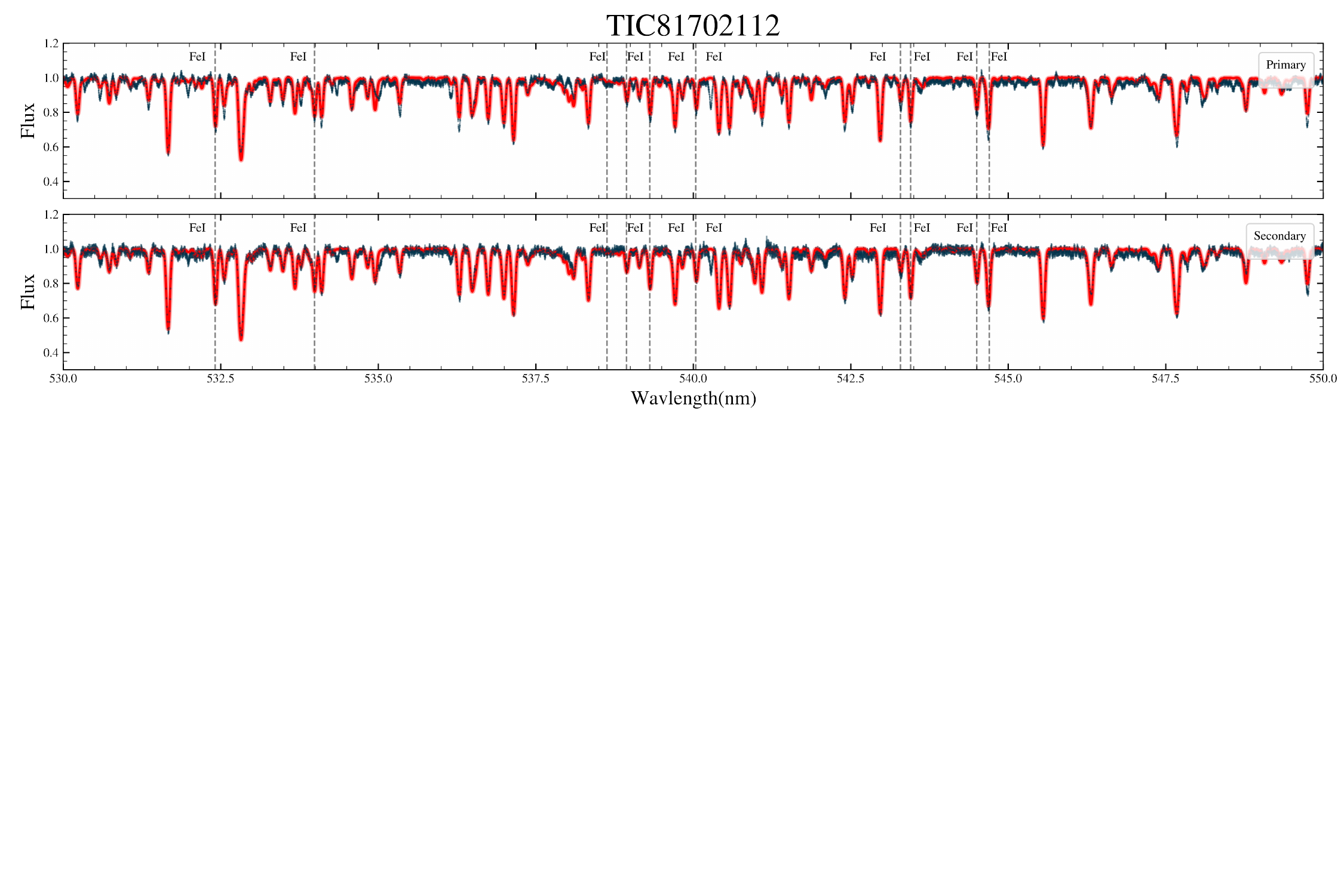}
    \label{fig:sub1}
    \vspace{-6.5 cm} 
  \end{subfigure}
  \begin{subfigure}[b]{0.925\linewidth}
    \includegraphics[width=\linewidth]{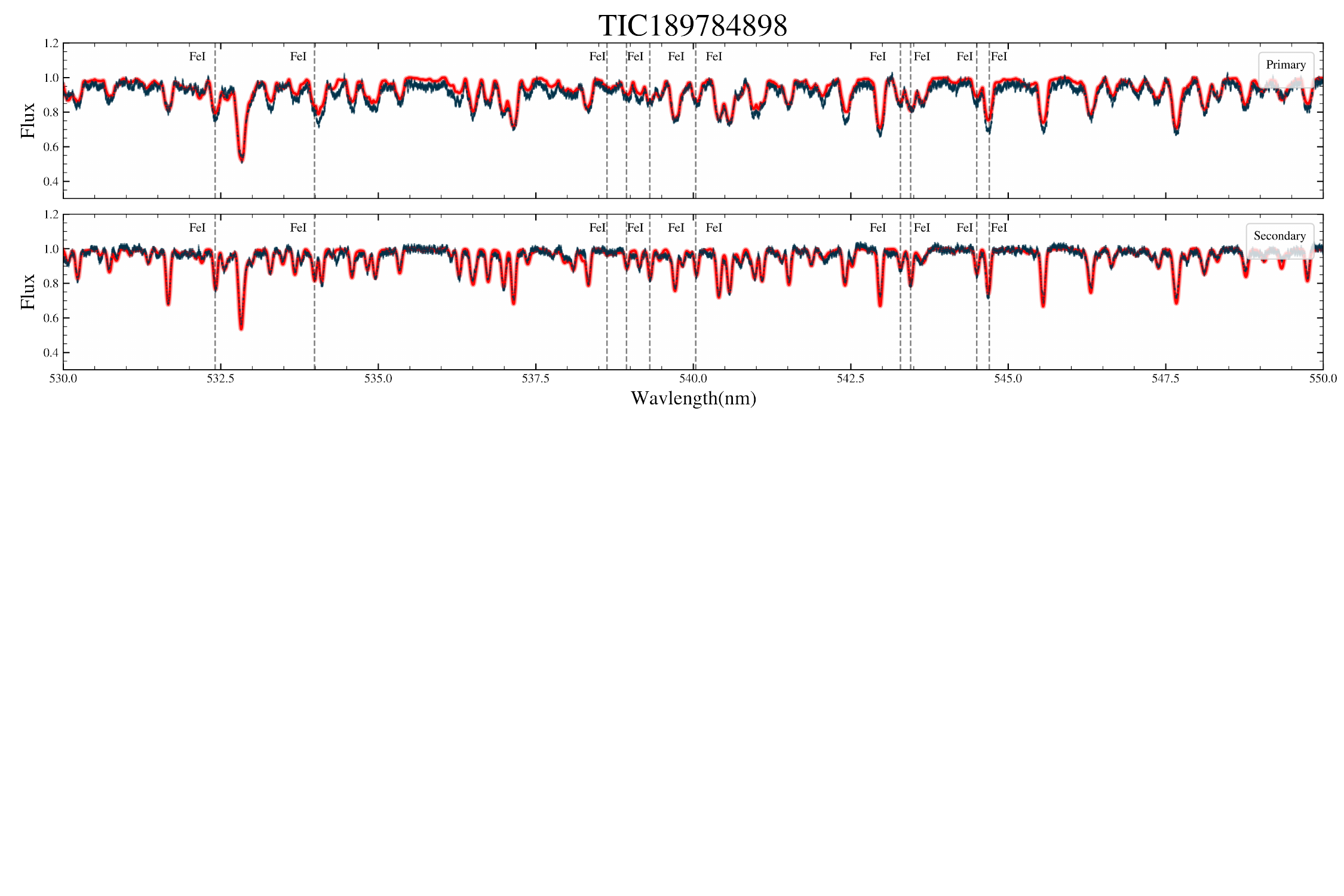}
    \label{fig:sub2}
    \vspace{-6.5 cm} 
  \end{subfigure}
    \begin{subfigure}[b]{0.925\linewidth}
    \includegraphics[width=\linewidth]{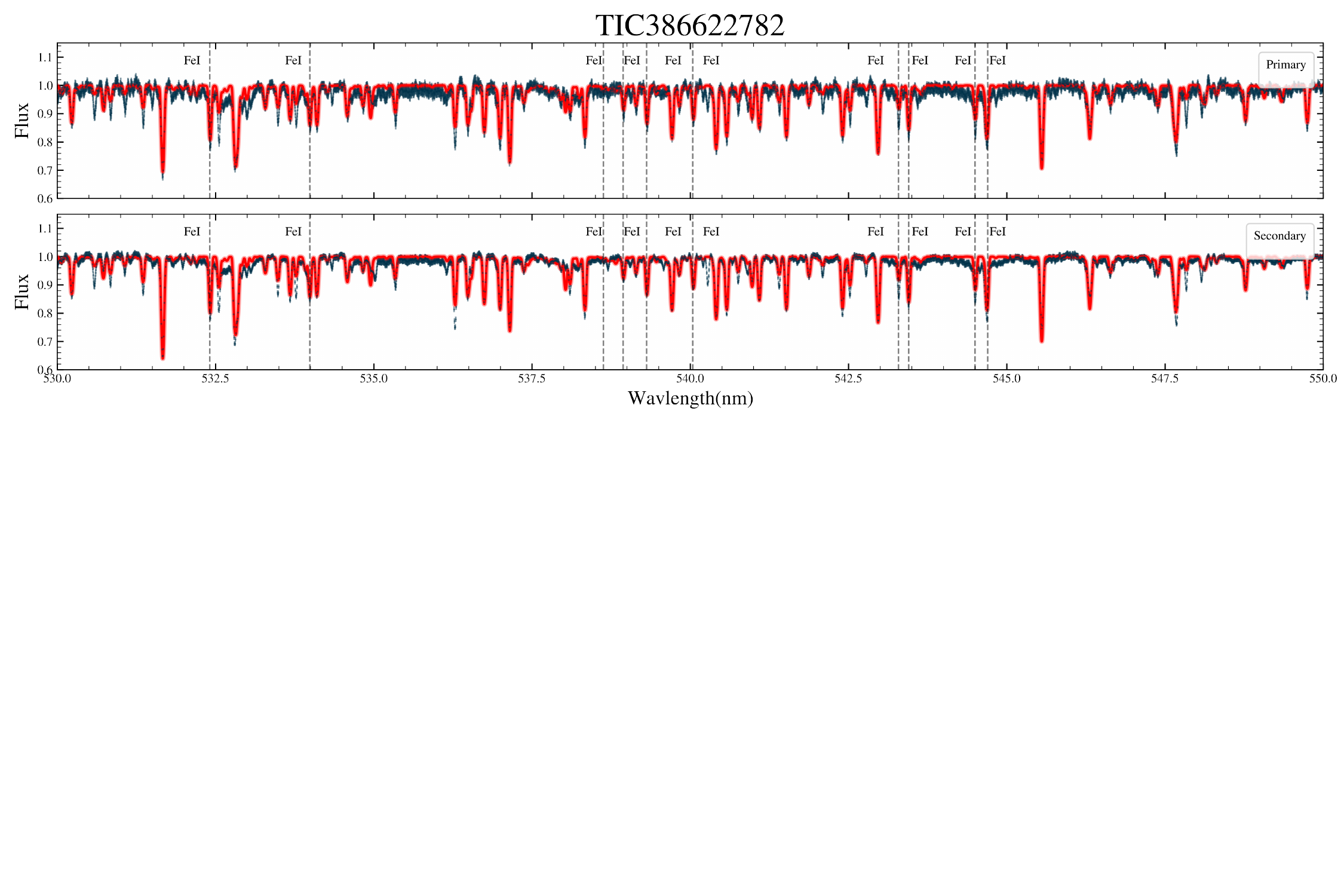}
    \label{fig:sub3}
    \vspace{-6.5 cm} 
  \end{subfigure}
  \begin{subfigure}[b]{0.925\linewidth}
    \includegraphics[width=\linewidth]{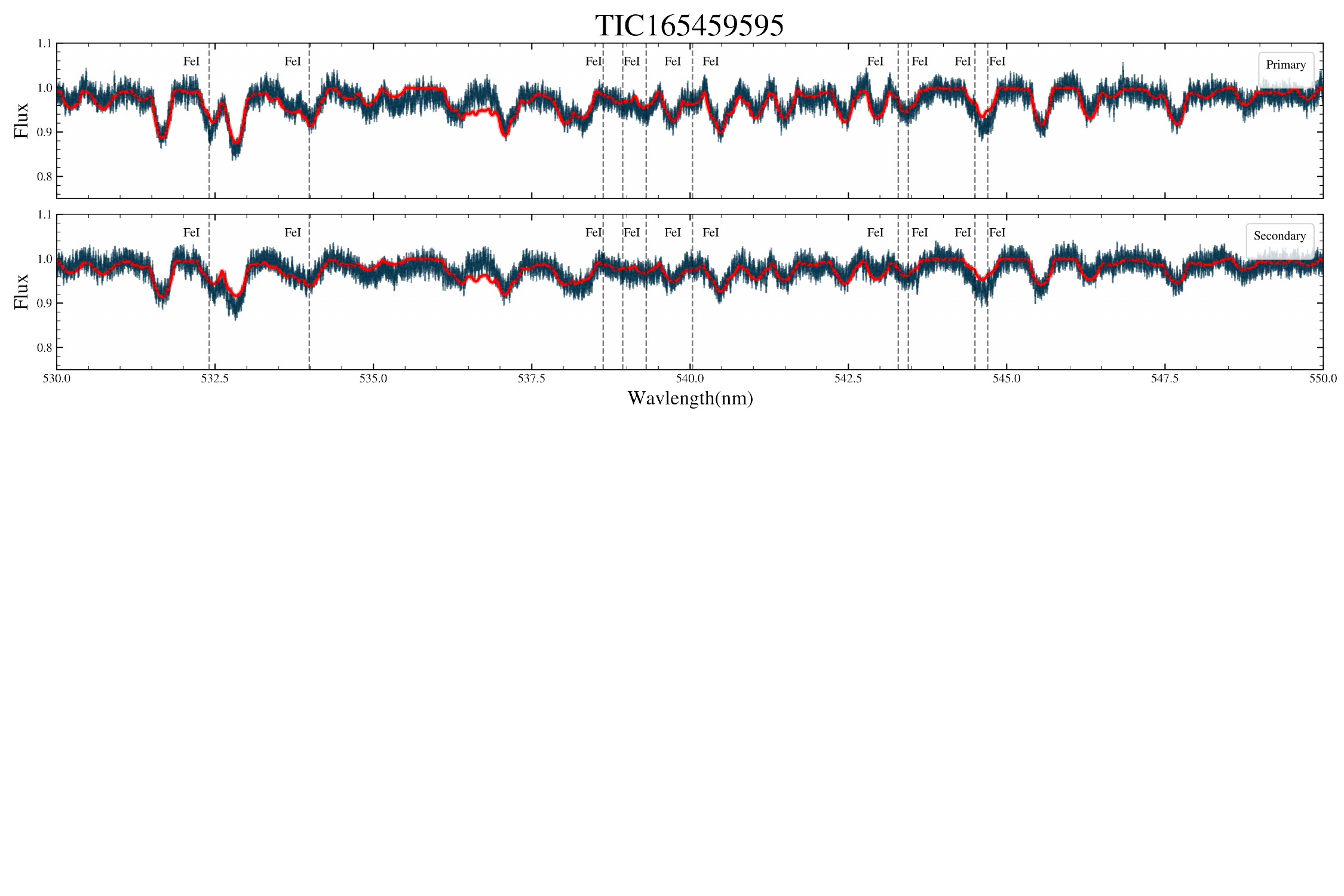}
    \label{fig:sub4}
    \vspace{-6.5 cm} 
  \end{subfigure}
  
  \caption{Model spectra (in red) overplotted on the disentangled spectra (in blue) of the primary and secondary components.}
  \label{fig:ispec_results}
\end{figure*}

During the fitting process, we fixed the spectral resolution to the known value, and the values of surface gravity ($\log(g)$) from the LC solution to minimise the degeneracy in fitted values. We performed the fitting procedure in two steps. Step one used the GES line-list\footnote{https://www.blancocuaresma.com/s/iSpec/manual/usage/synthesis} well suited to determine abundances in order to determine the global metallicity. In a trial run, we kept the $v\sin(i)$ free alongside other parameters to check if it roughly matched the value obtained from BF analysis. We then fixed it and estimated the metallicity. In this step, we excluded the line regions with broad-wing features (e.g. the Mg triplet), to make sure that fitted parameters were not affected by incorrect continuum normalisation. In the second step, we fixed the metallicity to the value obtained from step one, and using the line-list best suited for determining atmospheric parameters we fitted for $T\mathrm{_{eff}}$, $v\sin(i)$ and $v\mathrm{_{mic}}$. Macroturbulence velocity ($v\mathrm{_{mac}}$) was determined using an empirical relation as in \citet{2014A&A...566A..98B}, which is intended to be used for FGK-type stars ($\sim$ 4000-7000 K). Therefore, in the case of TIC~386622782 and TIC~165459595, we set it to zero to avoid degeneracy with $v\sin(i)$. The final parameter estimates obtained from this analysis are stated in Table~\ref{tab:ispec_results}, and the best-fit spectra are shown in Fig.~\ref{fig:ispec_results}.


\begin{table*}[htbp]
    \centering
    \small
    \caption{Atmospheric parameters obtained after analysis of the disentangled spectra using \textsc{ispec}.}
    \label{tab:ispec_results}
   \begin{tabular}{lcccc}
         Parameters &  TIC~81702112       & TIC~189784898      & TIC~386622782	   & TIC~165459595$^a$   \\
         \hline
$T_\mathrm{eff,A}$ &    6750 $\pm$ 70    &   5438 $\pm$ 210        &   7100 $\pm$ 300    &   8050 $\pm$ 350 \\
$T_\mathrm{eff,B}$ &    7030 $\pm$ 60    &   6300 $\pm$ 150        &   7800 $\pm$ 250    &   8325 $\pm$ 675  \\

$\log{g}_\mathrm{A}$ &   3.87 (fixed)    &  3.86 (fixed)           &   4.06 (fixed)	   &  4.36 (fixed)   \\
$\log{g}_\mathrm{B}$ &   3.70 (fixed)    &  3.45 (fixed)           &   3.86 (fixed)	   &  4.38 (fixed)    \\

$[M/H]_\mathrm{A}$ &   -0.08 $\pm$ 0.06  &   -0.11 $\pm$ 0.18       & -0.23 $\pm$ 0.18   &  - 	    \\
$[M/H]_\mathrm{B}$ &   -0.08 $\pm$ 0.04  &   -0.11 $\pm$ 0.12       & -0.23 $\pm$ 0.15   &  - 	    \\

$\alpha_\mathrm{A}$ &     0              &  0.02 $\pm$ 0.07         &    0		    &  -	    \\
$\alpha_\mathrm{B}$ &     0              &   0.04 $\pm$ 0.07        &    0		    &  -	     \\

$v_\mathrm{mic,A}$ &   3.02 $\pm$ 0.22   &  1.96 $\pm$ 0.18         &   1.1 $\pm$ 0.50     &   1.56 $\pm$ 0.76 \\
$v_\mathrm{mic,B}$ &   4.32 $\pm$ 0.26   &  2.09 $\pm$ 0.18         &   2.0 $\pm$ 0.25     &   1.58 $\pm$ 1.14 \\

$v_\mathrm{mac,A}$ &    12               &     3.71	            &      0 (fixed)  	   &   0 (fixed)    \\
$v_\mathrm{mac,B}$ &    15               &     3.80	            &      0 (fixed)  	   &   0 (fixed)     \\

$v_\mathrm{A}\sin{i_A}$ & 20 $\pm$ 1     &	47 $\pm$ 1          &      19 $\pm$ 3     &	90 $\pm$ 10  \\
$v_\mathrm{B}\sin{i_B}$ & 24 $\pm$ 1     &	28 $\pm$ 1          &      18 $\pm$ 2     &	113 $\pm$ 18  \\
\hline
\end{tabular}
\\ $^a$ $T\mathrm{_{effs}}$ obtained for TIC~165459595 have not been used for further analysis.

\end{table*}
\section{Light Curve modelling}
An expeditious look at the TESS LCs of our sample shows that they consist of well-separated stars. For the process of modelling the LCs, we needed something that was efficient but also robust and well-tested. We used version 40 of the \textsc{jktebop} code, \citep{2004MNRAS.351.1277S, 2004MNRAS.355..986S, 2013A&A...557A.119S} which itself is based on the \textsc{ebop} program \citep{1981AJ.....86..102P, 1981ASIC...69..111E}. Using the provided parameters, it computes the LCs through integration of concentric circles that approximate the surface of each star. It accounts for the oblateness of the stars by modelling them as bi-axial ellipsoids, and uses their projections to calculate proximity effects, which is a sufficient approximation for oblateness levels below $\sim$4\%. 
\par
\textsc{jktebop} adopts the relative radii $r\mathrm{_A}$ and $r\mathrm{_B}$, stellar radii relative to the orbital separation, as parameters governing the shape of the eclipses. This is implemented using two input values, $r\mathrm{_A}$/$r\mathrm{_B}$ and $r\mathrm{_A}$+$r\mathrm{_B}$. We set both as free parameters. Other parameters set free during the fitting process are the inclination ($i$), surface brightness ratio ($J$), eccentricity ($e$), and argument of periastron ($\omega$). This code also provides an option to fit for a third light ($l_3$). In our test runs, we found the values of $l_3$ to be either very close to zero or nonphysical in nature. Scatter arising from the pulsations is an added reason why we simply fixed the $l_3$ parameter to zero (in all cases except for TIC~165459595 due to its proximity to two other stars). The coefficients of gravitational and limb darkening were set values sourced from \citet{2017A&A...600A..30C}. This version of the code also allows the user to fit polynomials and (up to 9) sines to account for additional trends in the data affecting the binary LC. We used polynomial functions to account for some long-term trends and also sine functions in some cases to model possible signatures from a stellar spot.
\par
For cases in which the RVs are available, we used them within the code to perform a simultaneous LC-RV fit. This provides absolute stellar masses and radii as outputs instead of relative radii. For cases in which RV data has high uncertainties, we used the $K_1, K_2$ values obtained using \textsc{v2fit}, and \textsc{jktabsdim} \citep{2005A&A...429..645S} to calculate the absolute parameters of the system. \textsc{jktabsdim} was also used to obtain distance estimates for all the targets.
\par
\textsc{jktebop} provides a way to obtain reliable error estimates for the fitted parameters by implementing an in-built Monte Carlo (MC) algorithm. We exploited this feature and performed 25,000 MC runs for each target to obtain the parameter uncertainties. The corner plots for the key parameters are made available in Figure~\ref{fig:Corner_plots}. The correlations observed in these corner plots are well known, and are often showed in other studies of EBs \citep{2023MNRAS.521..677P}. The differences between each corner plot are likely due to the different character of the LC: some show a total eclipse, while others are only grazing, and sometimes eclipses are nearly identical and at other times they are very different, as in the case of TIC~386622782.\\


\section{Age and evolution}
Stellar mass is arguably the most crucial parameter determining the evolution of a star. Therefore, stellar mass and the corresponding radii obtained through LC modelling are good parameters to compare against the stellar isochrones. However, the mass-radius plane is degenerate to the combination of age-metallicity values, requiring additional constraints. This problem can be relieved by checking if the star is located at the same evolutionary stage in the mass-$T\mathrm{_{eff}}$ \citep{2019MNRAS.484..451H}.
\par
\par
In order to perform a simultaneous minimisation, we implemented a multi-parameter fitting procedure based on the Mahalanobis distance \citep{mahalanobis1936generalized}. Using the \textsc{ezmist}\footnote{https://github.com/mfouesneau/ezmist} package, we created a database of isochrones from MESA Isochrones and Stellar Tracks (MIST) \citep{2016ApJS..222....8D, 2016ApJ...823..102C}, from ages of 1 Myr to 10 Gyr in steps of 0.1 on a logarithmic scale. The range of metallicities from -0.25 to 0.25 was covered, and the initial abundance value was Z=0.0142. 
\par
For the fitting procedure, we created a covariance matrix for the combination of masses, radii, and $T\mathrm{_{eff}}$, independently for each star in the binary. We then calculated the Mahalanobis distance for this point in phase space to every row of the isochrone database. The minimum Mahalanobis distance, corresponding to the minimum $\chi^2$, was then used to visualise the isochrones around the corresponding age.
\par
Since the two stars in the binary are being fit independently within this procedure, it is possible to obtain different ages for both stars. The assumption of co-evolution is then to be applied by the user to choose from the best-fit ages around the obtained results for the two stars.
\par
We tested this procedure on two systems, KIC~7821010 and KIC~9641031, from the \citet{2019MNRAS.484..451H} study, to validate its effectiveness. The age estimates for these targets from the study were reproduced within 2-$\sigma$ bounds around the parameters. The procedure is especially useful due to its flexibility with respect to the number of stars and the number of parameters that can be used for each of the stars. Within the scope of this paper, we did not compute any statistical errors within the fitting routine. Instead, we derived isochrones corresponding to the parameter uncertainty limits and used these to define the uncertainty range for the age estimates.
\par

\section{Results and discussions}
In this section, we present the results from our analyses for the four new candidate $\delta$ Scuti pulsators in EBs, and also revised parameters for TIC~308953703 using TESS photometry.

\subsection{TIC~81702112 = HD 97329}
TIC~81702112 was classified as a variable star in the All Sky Automated Survey (ASAS) catalogue of variable stars \citep{2002AcA....52..397P} as ASAS~J111134-4956.2. A preliminary LC+RV solution, based on ASAS photometry and RV measurements of a precision lower than ours, has been presented in \citet{2016PASP..128g4201K}. The system was revisited by the TESS, obtaining 2-min cadence photometry in sectors 10, 63, and 64. A brief look at the LC reveals that at least one component of this EB is pulsating.
\par
An orbital solution for this target was obtained using ten CHIRON spectra.  We initialised the \textsc{jktebop} model using $q$, $e$, and $\omega$ obtained using the \textsc{v2fit}, attempting a simultaneous LC-RV fit. However, after several unsuccessful attempts, we realised that the RVs of the secondary star with large $rms$ may be responsible for the erroneous fitting outcomes. This large $rms$ may be due to the high-amplitude pulsations, hinting that the secondary star may be the pulsator. We then used RVs only from the primary component and a good fit was obtained. We verified that the parameters for this single-component RV fit were consistent with the ones from \textsc{v2fit}.
\par
\begin{figure}[htbp]
         \centering
         \includegraphics[width=0.9\columnwidth]{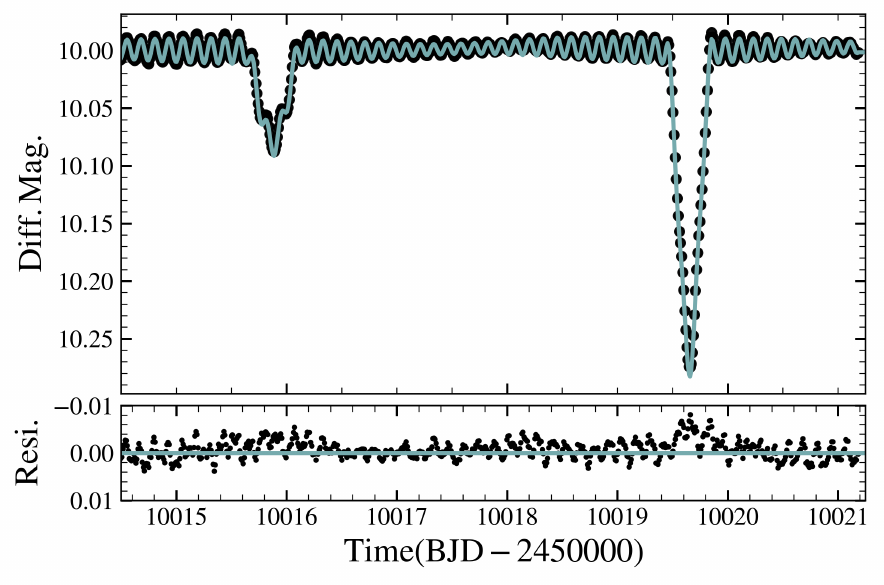}
         \caption{120-s cadence TESS LC of TIC~81702112 (filled circles), overplotted with the best-fit \textsc{jktebop} model. The variable amplitude of pulsations was fitted using a combination of sines.}
         \label{fig:A111134_sine_model}
\end{figure}

Due to the high-amplitude pulsations affecting the eclipse profiles, and hence the estimation of stellar radii, it was necessary to account for them during the modelling. We extracted the dominant pulsation frequencies using \textsc{period04} \citep{2005CoAst.146...53L}, and employed a total of three sine waves using these fixed frequencies. The starting epoch and amplitude of these sine waves were kept variable along with ($J$), ($i$), ($r\mathrm{_A}$+$r\mathrm{_B}$), ($r\mathrm{_A}$/$r\mathrm{_B}$), the time of superconjunction ($t\mathrm{_o}$), and the orbital period ($P_\mathrm{orb}$). The best-fit model obtained after following this treatment is shown in Fig.~\ref{fig:A111134_sine_model}.

Using the best-fit parameters from the final model with pulsations, we made another model without the addition of sinusoidal functions. This \enquote{clean} eclipsing model was then removed from the LC, and the residuals were used for pulsation analysis. This model is shown in Fig.~\ref{fig:A111134_lc_model}.


\par
Through multiple attempts made for the error estimation via MC runs, it was clear that reliable parameters were obtained when sines were simultaneously fitted to account for higher-amplitude pulsations, highlighting the utility of this feature in \textsc{jktebop}. The errors resulting from the MC runs translate to 1.8\% and 1.4\% errors on the absolute radii of the primary and secondary, respectively. The orbit is eccentric, which may have a role to play in the variability of the amplitude of the pulsation frequency.

\par
The spectroscopic analysis in \textsc{iSpec} confirms that the stars have almost identical [M/H], very similar $T\mathrm{_{eff}}$, and $v\sin(i)$. Using the parameter estimates from EB modelling and \textsc{iSpec}, we determine the age of this system to be $\sim$1.125~Gyr, as is displayed in Fig~\ref{fig:A111134_age}. Both components seem to be at the MS stage; however, the more massive one is at the very end of this phase. The estimated distance lands around 303$\pm$7~pc, which is in excellent agreement with the GDR3 value of 300$\pm$2~pc. This is supporting evidence towards validation of the obtained radii and $T_\mathrm{eff}$. An E(B-V) of 0.005 was assumed to get a consistent distance between different photometric bands within \textsc{jktabsdim}.

The synchronisation velocities given by \textsc{jktabsdim}, for the less massive and more massive components, are $\sim$ 17 and 21 km/s, respectively. These values are valid for circular orbits. In the case of eccentric orbits, a pseudo-synchronisation is achieved when the ratio of orbital period and rotation period equals

\begin{equation}
\centering
    \frac{P\mathrm{_{orb}}}{P\mathrm{_{rot,ps}}} = \frac{1+7.5e^2+5.625e^4+0.3125e^6}{(1-e^2)^{3/2}(1+3e^2+0.375e^4)}
\label{psv}
\end{equation}

Using Eq.\ref{psv} \citep{1981A&A....99..126H,2008EAS....29....1M}, we derived the pseudo-synchronisation velocities ($v\mathrm{_{rot,ps}}$) for the less massive and more massive star to be $\sim$ 21 and 26 km/s, respectively. The values are in 2-$\sigma$ agreement with the $v\sin(i)$ values for these components obtained from the spectral analysis.


\subsection{TIC 189784898 = V Cir}
The classification of TIC 189784898 as a variable star dates back to 1932 \citep{1932VeBB....9D...1G}. This variability was later accredited to eclipses \citep{1967IBVS..208....1S, 2012A&A...548A..79A}. More recently, \citet{2010NewA...15..150Z} presented LC analysis of data obtained with the INTEGRAL-OMC instrument. The target was revisited by the TESS and 2-min cadence photometry was obtained in sector 38.

We initiated the simultaneous LC-RV modelling using the orbital parameters from \textsc{v2fit}. The LC demanded a single polynomial together with a total of four sine functions alongside the binary model to account for trends likely arising due to stellar spot(s). These were fixed to $P_\mathrm{{orb}}$, $P_\mathrm{{orb}}$/2, $P_\mathrm{{orb}}$/3, and $P_\mathrm{{orb}}$/4, while keeping the starting epoch and amplitude free. Initially, we attempted to fit two sines per component to check their contribution towards the final fit; however, we found that spots on the primary component were preferred in order to obtain the best fit. Due to the lower number of RVs and variability near eclipses due to spots, we chose to fix $e$ and $\omega$ to zero. The best-fit obtained following this process is shown in Fig.~\ref{fig:VCir_lc_model}.
\par
The \textsc{iSpec} analysis reveals a hotter primary that is rotating more slowly than the larger secondary component. The metallicity estimates of the two stars match. It is slightly lower than the solar value; however, the errors are large enough to prevent us from saying this with absolute certainty. The high S/N disentangled spectra also allowed us to check the spectroscopic $\log(g)$, which agreed with the photometric $\log(g)$ within errors for both the components.
\par
The distance calculated using \textsc{jktabsdim}, 575$\pm$23~pc, agrees with the GDR3 distance of 549$\pm$5~pc. To achieve a consistent distance between different bands within \textsc{jktabsdim}, the $E(B-V)$ of 0.2 needed to be used. The age for this system is estimated to be $\sim$1.75~Gyr. As is shown in Fig.~\ref{fig:VCir_age}, the more massive component has already evolved into a sub-giant, and the less massive star is at the end of its MS.


\subsection{TIC 386622782 = HR 2214}
The discovery and classification of TIC~386622782 as a \enquote{double star} dates back to 1950 \citep{1950JO.....33...93M, 1950AJ.....55..153W}. The system is a visual binary, with one of the components being an SB2. This object has been observed over multiple campaigns \citep{1993yCat.3135....0C, 2002A&A...393..897R, 2004A&A...424..727P, 2014ApJ...791...58A} in attempt to constrain the spectral type and the rotational and radial velocities. It is described across the literature as a chemically peculiar star \citep{1959ApJ...129...88S,1985ApJS...59..229A,2009A&A...498..961R,2018MNRAS.480.2953G}. The eclipsing (and pulsating) nature of this system was first noted during an inspection of the TESS data by M.~Pyatnytskyy in January 2022, and reported to the International Variable Star Index (PMAK~V115\footnote{\url{https://www.aavso.org/vsx/index.php?view=detail.top&oid=2225979}}). 

\par
A first look at its TESS LC shows primary eclipses at a temporal separation of 23.80841 days, and a complex pulsation signal containing multiple frequencies. A careful inspection of the phase-folded LC reveals a tiny secondary eclipse at a phase of $\sim$0.49. Modelling the tiny secondary eclipse in \textsc{jktebop} was challenging because the amplitude of pulsations is of the same order as that of the eclipse depth. Fixing the value of $e$ and $\omega$ obtained from \textsc{v2fit}, we performed a simultaneous LC-RV fit. This was necessary to make sure that the model does not \enquote{miss} the tiny secondary eclipse. We primarily aimed to constrain the $J$, $i$, $r\mathrm{_A}$, and $r\mathrm{_B}$. A total of six polynomials were used to offset the trends affecting different sections of the LC. The best-fit model, as is seen in Fig.~\ref{fig:HR2214_lc_model}, is able to reproduce the tiny secondary eclipse. The geometry of this eccentric system gives rise to such a small secondary eclipse in spite of this star contributing more than 80\% of the total flux in the observed band. In this case, the availability of a large number of RV measurements played a crucial role in obtaining a reliable binary model. The distribution of the parameters during MC runs is displayed in a corner plot in Fig.~\ref{fig:HR2214_mc_model}.


The metallicity values for the two components are consistent with each other. Using the obtained $T\mathrm{_{eff}}$ estimates, the distance to this system was found to be $103 \pm 3$ pc. This value lies just within the 3-$\sigma$ range of the last updated value from the Hipparcos catalogue \citep{2007A&A...474..653V}. We found that an isochrone of age $\sim$0.8 Gyr reproduced the two stars in fair agreement with the parameters obtained from our analysis. However, the temperature of the less massive component is nearly 4-$\sigma$ away from the corresponding value on the isochrone. This isochrone fit is shown in Fig. \ref{fig:HR2214_age}.

\subsection{TIC 165459595 = V1109 Cen}
During our sample selection from the CR\'EME database, TIC~165459595 was included as one of the components was near the 2.5 $M_\mathrm{{sun}}$ upper limit. While collecting further data on the sample, it was noted that TIC~165459595 has recently been flagged as one of the many EBs observed by TESS that exhibit pulsations \citep{2022ApJS..259...50S, 2022ApJS..263...34C}. A detailed exploration of its physical as well as oscillatory nature is therefore timely. 
\par
An orbital solution for this target was obtained using six CHIRON spectra. The unsatisfactory quantity of spectra and high rotational velocities of both components hampered the calculation of precise $K_1, K_2$, resulting in much larger errors on masses compared to the other targets in the sample. We initialised the \textsc{jktebop} model using $q$, $e$, and $\omega$ obtained from \textsc{v2fit}, attempting a simultaneous LC-RV fit using the 2-min cadence TESS photometry from sector 37. The MC errors of $\sim$0.3\% obtained on the relative radii translated to $\sim$1.4\% errors on the absolute radii of the stars. The best-fit model is shown in Fig.~\ref{fig:V1109_Cen_lc_model}

\par
In spite of the noisy nature of composite spectra and the low number of observations, we tried disentangling the components and modelling them within \textsc{iSpec}. In this case we fixed all of the parameters, except the $T\mathrm{_{eff}}$ and $v_\mathrm{{mic}}$. After a few unsuccessful attempts, we chose not to fit for the [M/H]. The two-fold reasons for erroneous results could be high rotation velocity and relatively low S/N of the decomposed individual spectra. To determine the $T\mathrm{_{eff}}$, the [M/H] was set to the GDR3 estimate of -0.44. The bottom panel of Figure \ref{fig:ispec_results} shows that the model obtained using ispec (red) does not fit very well, particularly in the 535-538 nm section. This could be due to poor S/N or incorrect continuum normalisation or both.

The process of determining the age for this system was challenging due to the unavailability of reliable spectroscopic temperature constraints. We attempted to constrain it using the flux ratio obtained from the EB modelling. We did this using \textsc{isofitter}, explained in further detail in \citet{2023MNRAS.521.1908M}, which searches through a grid of MIST isochrones and finds the best fit for the given set of parameters. Keeping the $l_3$ parameter free results in a $L\mathrm{_B}/L\mathrm{_A}$ value of $\sim$ 0.9, meaning the less massive primary star is brighter. This corresponds to an age of $\sim$2.37 Myr. However, when $l_3$ was kept fixed to zero, we obtained an $L\mathrm{_B}/L\mathrm{_A}$ that is slightly greater than 1. The difference between the residuals obtained after subtracting these two models from the LC is negligible. This prefers the slightly older age of $\sim$100 Myr. The errors in the mass estimates for the stars are quite large, making it very difficult to better constrain the age. As is shown in Fig.~\ref{fig:V1109_age}, the best-fit isochrone represents stars with temperatures around 13000 K.



\par
The calculation of distance for the other targets, within \textsc{jktabsdim}, used the \citet{2004A&A...426..297K} relations. However, these are valid for temperatures up to 10,000 K. The distance calculated using the relations of \citet{2002A&A...391..195G} is $\sim$ 400 pc and does not agree with the GDR3 distance of 310 $\pm$ 2. Also, consistency between the different photometric bands is obtained only if the E(B-V) is set to $\sim$ 0.45.

\subsection{TIC$~$308953703 = HIP~7666}
TIC$~$308953703 has been revisited multiple times over the past 40 years \citep{1985A&AS...59..461O, 1993yCat.3135....0C, 2017MNRAS.465.1181L} and is classified as an EB with one of the components exhibiting pulsations \citep{2005A&A...434.1063E}. Recently, \citet{2021MNRAS.508..529F} performed an extensive multi-site photometry and spectroscopic study to derive the absolute parameters of the binary, and study the pulsations in further detail. Three significant pulsation frequencies were identified and mode identification was attempted. However, the authors suggested that photometry from space missions is needed to improve the reliability of these mode identification results.

This target now has a 2-min cadence TESS (sector 58) photometry. We truncated the LC to four orbits to optimise computational time, while obtaining a good resolution between pulsation frequencies. 
We initialised our LC model together with the RV data obtained by \citet{2021MNRAS.508..529F}. During the initialisation of our model, we identified a typographical error in one of the RV timestamps provided. Specifically, the value 329.75669 was incorrectly recorded as 331.75669. The RV data has insufficient phase coverage, which adds the possibility of a small but measurable value of $e$. To check this, we set $e$ as a free parameter, at a fixed omega of $90^{\circ}$ (expected from the apparent phase of the secondary minimum). Due to the lower amplitude of pulsation frequencies, it was possible to obtain a good fit without implementing additional sines during the \textsc{jktebop} runs, except one for the out-of-eclipse ellipsoidal variability.

\begin{figure*}[h!]
\centering
\begin{subfigure}{0.49\textwidth}
\centering
  \includegraphics[width=\textwidth]{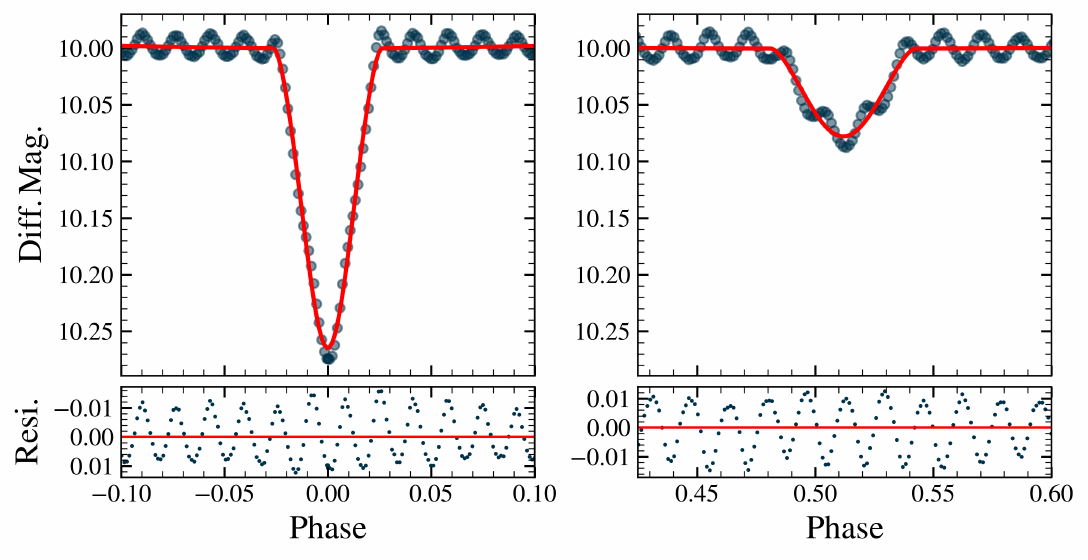}
  \caption{TIC~81702112}
  \label{fig:A111134_lc_model}
\end{subfigure}
\hfill
\begin{subfigure}{0.49\textwidth}
\centering
  \includegraphics[width=\textwidth]{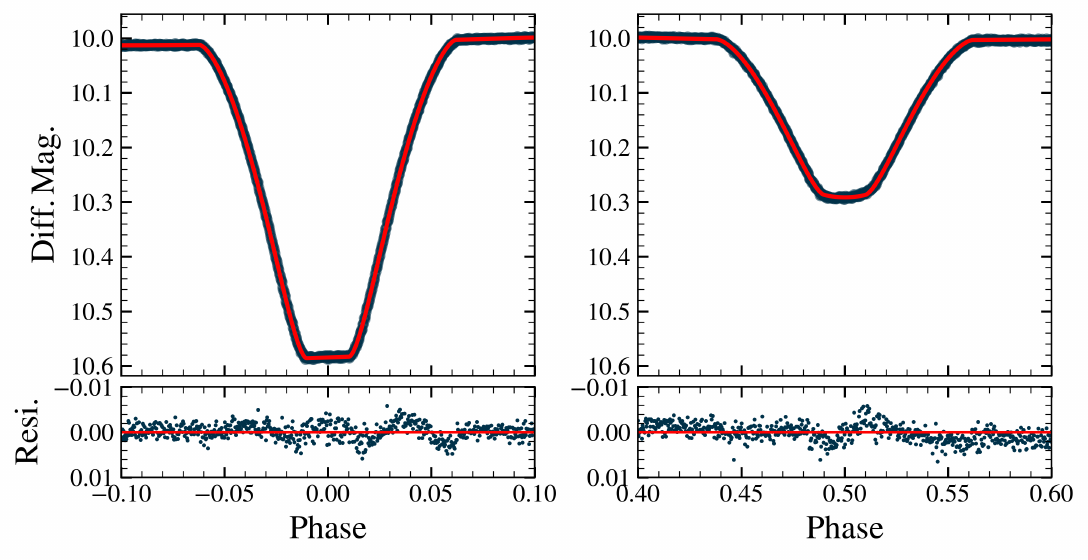}
  \caption{TIC~189784898}
  \label{fig:VCir_lc_model}
\end{subfigure}
\\
\begin{subfigure}{0.49\textwidth}
\centering
  \includegraphics[width=\textwidth]{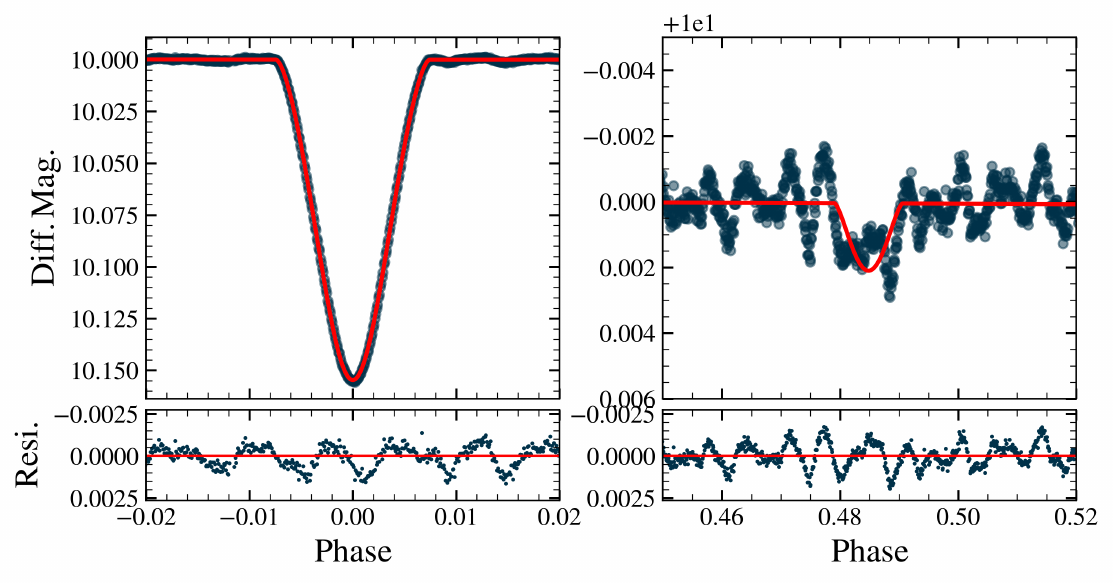}
  \caption{TIC~386622782}
  \label{fig:HR2214_lc_model}
\end{subfigure}
\hfill
\begin{subfigure}{0.49\textwidth}
\centering
  \includegraphics[width=\textwidth]{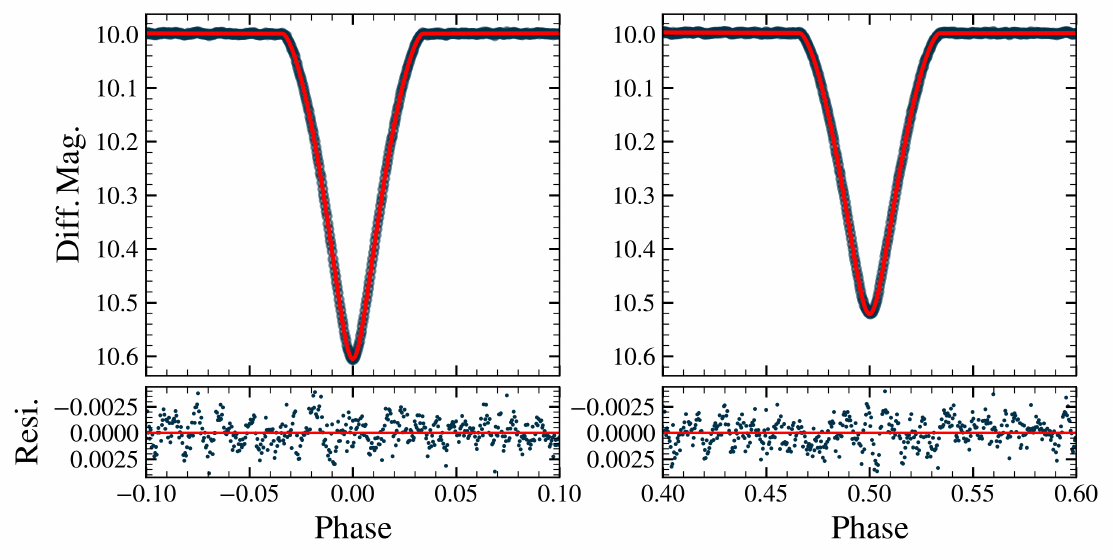}
  \caption{TIC~165459595}
  \label{fig:V1109_Cen_lc_model}
\end{subfigure}
\\
\begin{subfigure}{0.49\textwidth}
\centering
  \includegraphics[width=\textwidth]{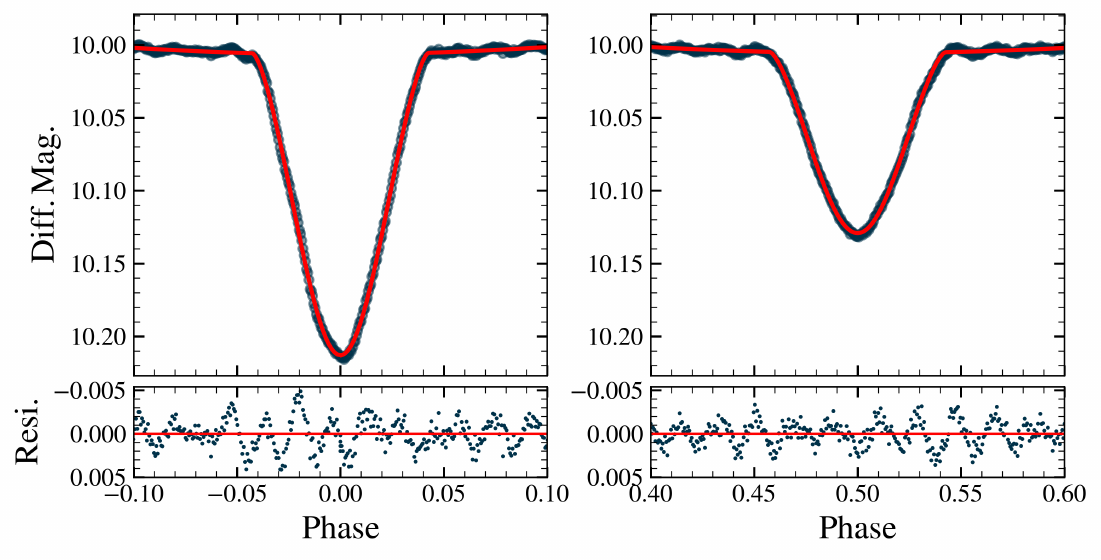}
  \caption{TIC$~$308953703}
  \label{fig:HIP7666_lc_model}
\end{subfigure}

\caption{\textsc{jktebop} binary models without using sines to account for pulsations. On the left panel is the primary eclipse and on the right is the secondary eclipse, with residuals plotted in the panel below. The error bars represent 3-$\sigma$ regions around the mean parameter values.}
\end{figure*}

\begin{figure*}[htbp]
\centering
\begin{subfigure}{0.33\textwidth}
\centering
  \includegraphics[width=\textwidth]{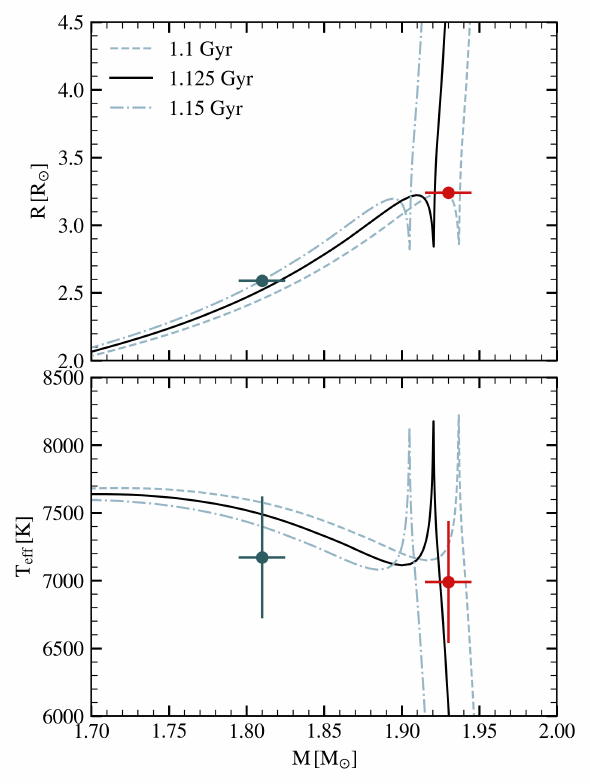}
  \caption{TIC~81702112}
  \label{fig:A111134_age}
\end{subfigure}
\hfill
\begin{subfigure}{0.32\textwidth}
\centering
  \includegraphics[width=\textwidth]{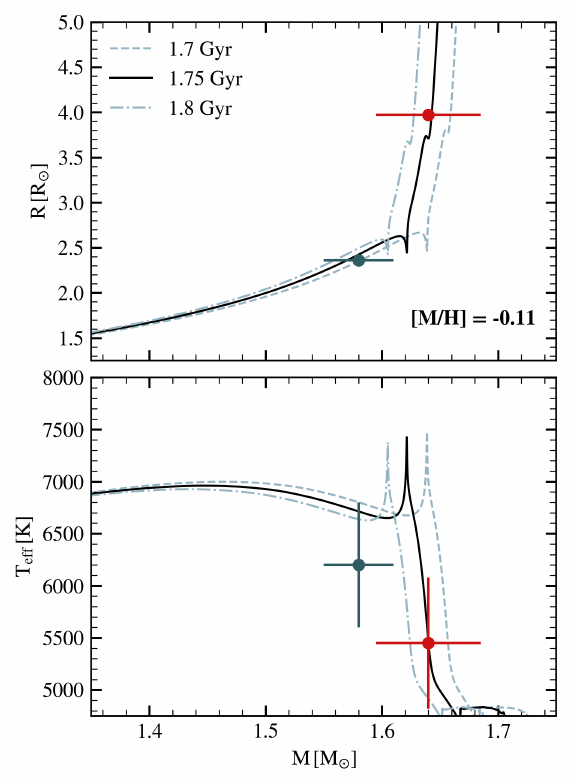}
  \caption{TIC~189784898}
  \label{fig:VCir_age}
\end{subfigure}
\hfill
\begin{subfigure}{0.32\textwidth}
\centering
  \includegraphics[width=\textwidth]{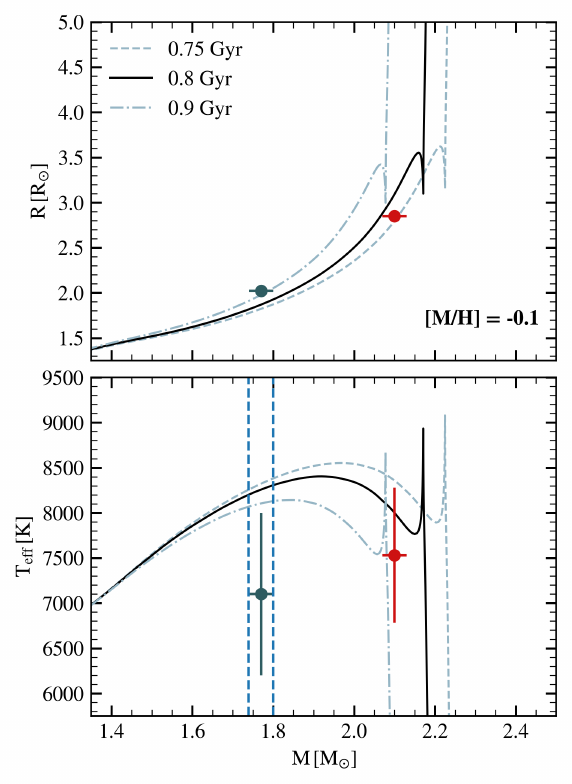}
  \caption{TIC~386622782}
  \label{fig:HR2214_age}
\end{subfigure}
\\
\begin{subfigure}{0.32\textwidth}
\centering
  \includegraphics[width=\textwidth]{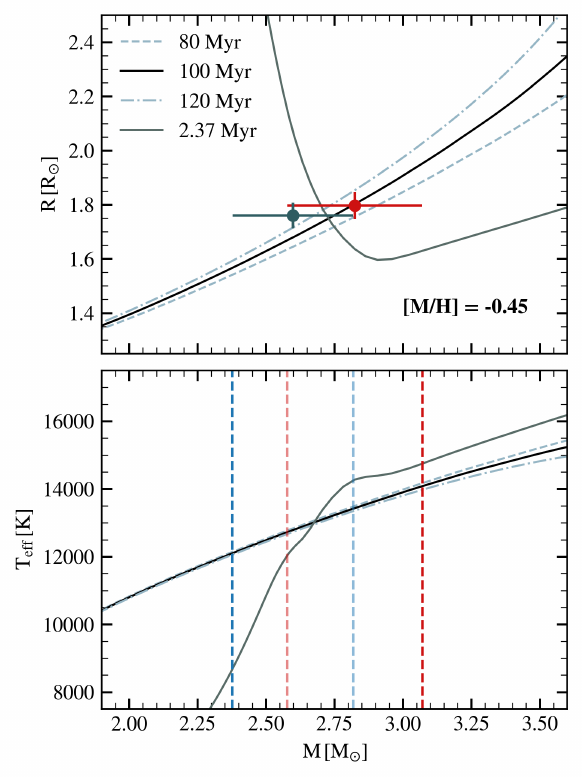}
  \caption{TIC~165459595}
  \label{fig:V1109_age}
\end{subfigure}
\begin{subfigure}{0.32\textwidth}
\centering
  \includegraphics[width=\textwidth]{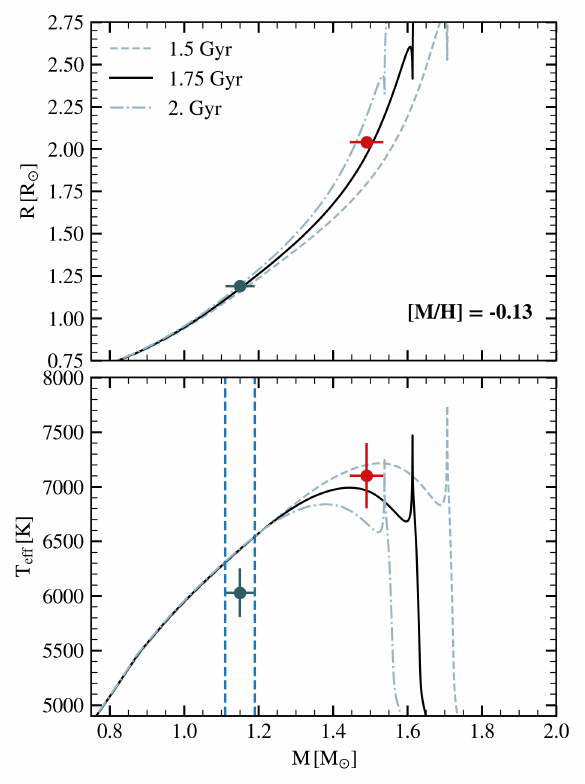}
  \caption{TIC$~$308953703}
  \label{fig:HIP7666_age}
\end{subfigure}

\caption{Results from the isochrone fitting routine, indicating the positions of the stars in the mass-radius (top) and mass-($T\mathrm{_{eff}}$) (bottom) diagrams. The black line indicates the best age estimate obtained for the set of input stellar parameters. The filled red and blue circles indicate the more and less massive components, respectively. In the case of TIC~165459595, the error bars represent the 2-$\sigma$ region instead of 3-$\sigma$.}
\end{figure*}

A good fit is obtained at $e=0.02697$, and it is shown in Fig.~\ref{fig:HIP7666_lc_model}. This resulted in slightly different values of masses and radii compared to the \citet{2021MNRAS.508..529F} results. These updated parameters were used to determine the age of the system. \citet{2021MNRAS.508..529F} discussed the inconsistency between the parameter estimates for the cooler secondary star and the one from the MESA evolutionary tracks. As is shown in Fig.~\ref{fig:HIP7666_age}, the updated masses and radii for the primary and secondary can now be represented by a single isochrone of age $\sim$1.75 Gyr, where both the stars are on the MS. 


We used the temperature estimates together with the updated parameters to estimate the distance to this system. The value provided by \textsc{jktabsdim} for the J-band --- 292$\pm$7~pc --- is consistent within 3-$\sigma$ of the GDR3 value of 313$\pm$1.3.

\section{Pulsation analysis}
To determine the pulsation frequencies for each system, we used the residuals left after subtracting the binary model from the LC. This ensures that effects arising due to ellipsoidal deformation and stellar spots are also removed to a good extent. We used v.1.2.9 of the \textsc{Period04}, based on \citet{2005CoAst.146...53L}, to perform the time series analysis. A typical pre-whitening procedure was followed to extract all the frequencies. The S/N ratio for detection, after pre-whitening for the previous frequencies, was kept at 5. All the significant frequencies were then simultaneously fitted using sinusoids in order to obtain their amplitudes and phases. A list of all the extracted frequencies, with their analytical errors \citep{1999DSSN...13...28M}, for each of the systems is provided in Appendix~\ref{tab:A111134_freqs}. Uncertainties on amplitude and phase were calculated using an MC simulation within \textsc{Period04}, keeping the frequency fixed to uncouple frequency and phase uncertainties.

\subsection{TIC~81702112}
\begin{figure}[h!]
         \centering
         \includegraphics[width=0.475\textwidth]{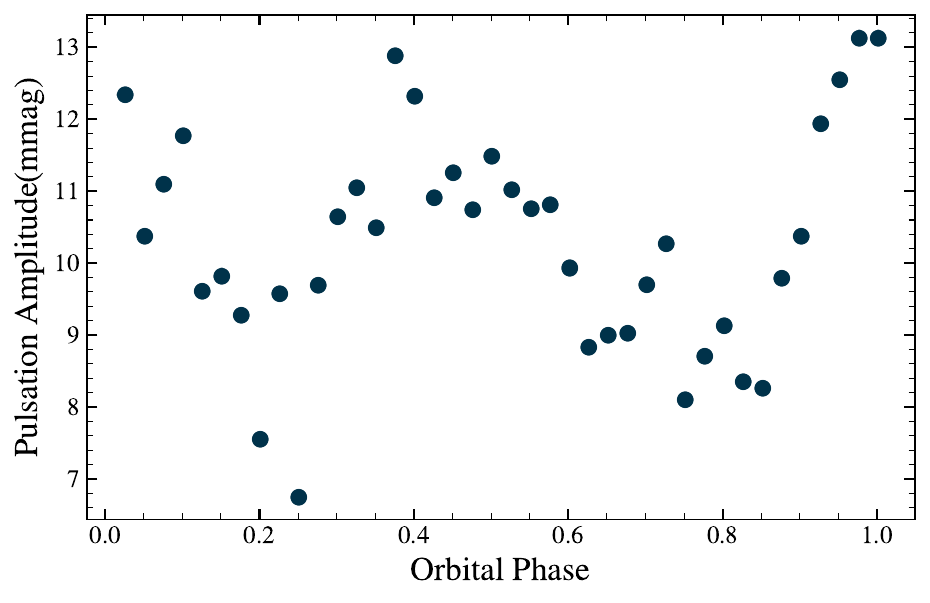}
         \caption{Amplitude of frequency with the highest power plotted over the orbital phase.}
         \label{fig:TIC~8170 amp var}
\end{figure}

The strongest pulsation frequency has an amplitude of $\sim$10~mmag (see Fig.~\ref{fig:TIC~8170_periodogram}), which is variable over the orbital phase. Such variability in the amplitude of pulsation frequency has been discussed in a few recently discovered systems \citep{2020NatAs...4..684H, 2020MNRAS.498.5730F}. The extent of this amplitude modulation over the orbital phase can be seen in Fig. \ref{fig:TIC~8170 amp var}.


\begin{figure}[h!]
         \centering
         \includegraphics[width=0.475\textwidth]{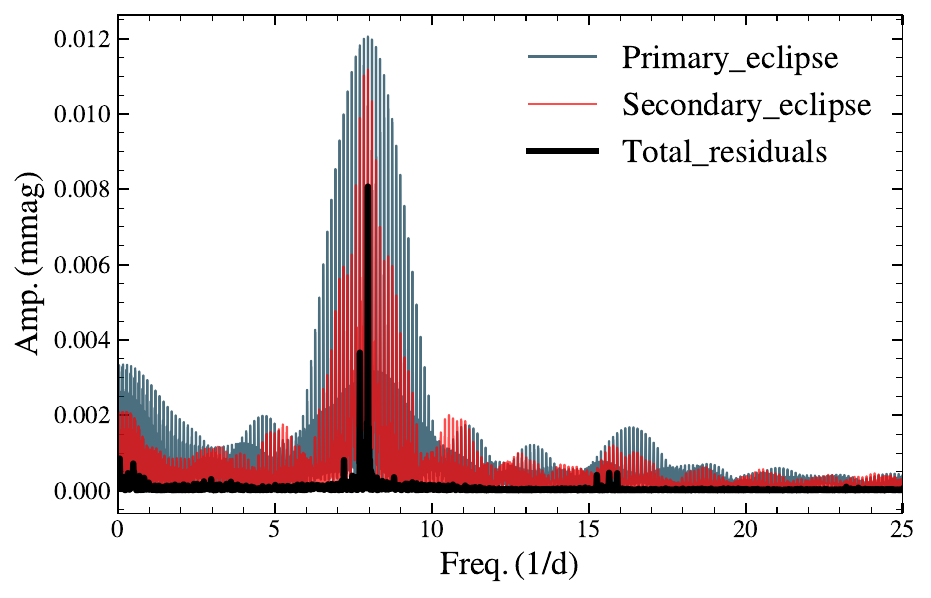}
         \caption{Frequency power spectra for primary (blue) and secondary eclipses (red) of TIC~81702112. Overplotted in black is the frequency spectra of the out-of-eclipse portions of the LC.}
         \label{fig:TIC~8170 ecl map}
\end{figure}

Suffering from the lack of total eclipses, it is difficult to state with confidence which of the two stars is pulsating. However, to get an idea, we followed the technique of analyzing the Fourier spectra in the primary and secondary eclipses. We chopped the residuals from the binary modelling to obtain these eclipse snippets, and overplotted their periodograms on top of each other as is shown in Figure~\ref{fig:TIC~8170 ecl map}. The amplitude of pulsations in the primary eclipse is slightly larger compared to the ones in the secondary eclipse. This is not so evident when visually inspecting the eclipses because of the difference in depths. The difference in pulsation amplitudes during the eclipses, though not large, hints towards the secondary being the pulsator. 

\subsection{TIC~189784898}
\begin{figure*}[htbp]
\centering
\begin{subfigure}{0.49\textwidth}
  \centering
  \includegraphics[width=\linewidth]{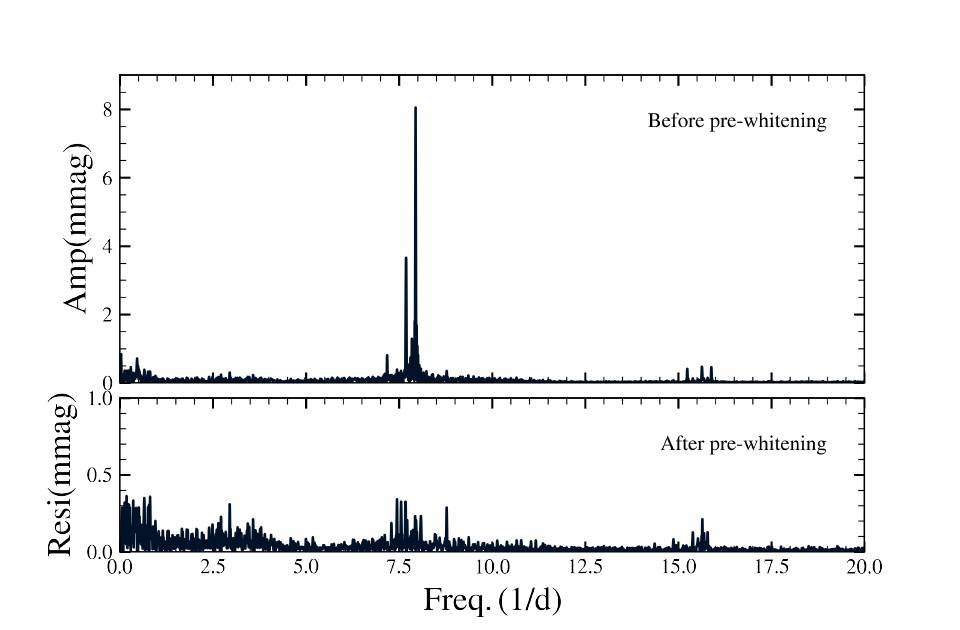}
  \caption{TIC~81702112}
  \label{fig:TIC~8170_periodogram}
\end{subfigure}
\hfill
\begin{subfigure}{0.49\textwidth}
  \centering
  \includegraphics[width=\linewidth]{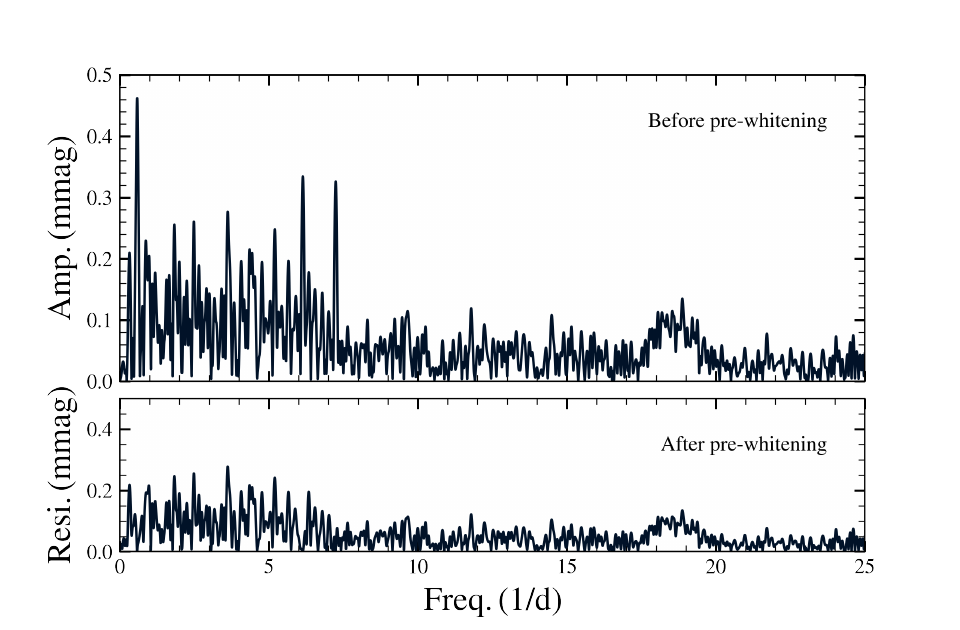}
  \caption{TIC~189784898}
  \label{fig:TIC~18978_period}
\end{subfigure}
\centering
\begin{subfigure}{0.49\textwidth}
  \centering
  \includegraphics[width=\linewidth]{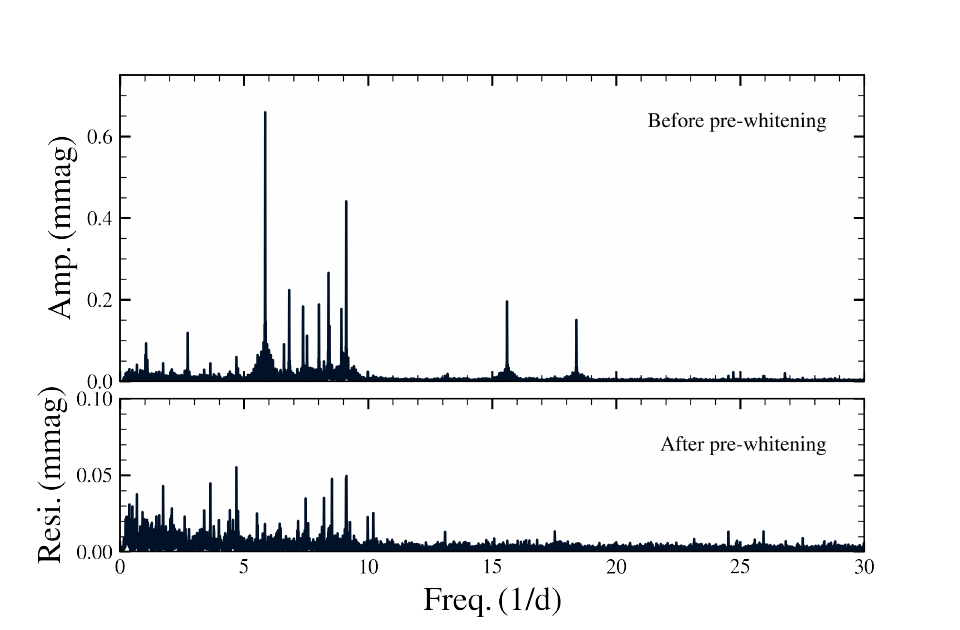}
  \caption{TIC~386622782}
  \label{fig:TIC~3866_period}
\end{subfigure}
\hfill
\begin{subfigure}{0.49\textwidth}
  \centering
  \includegraphics[width=\linewidth]{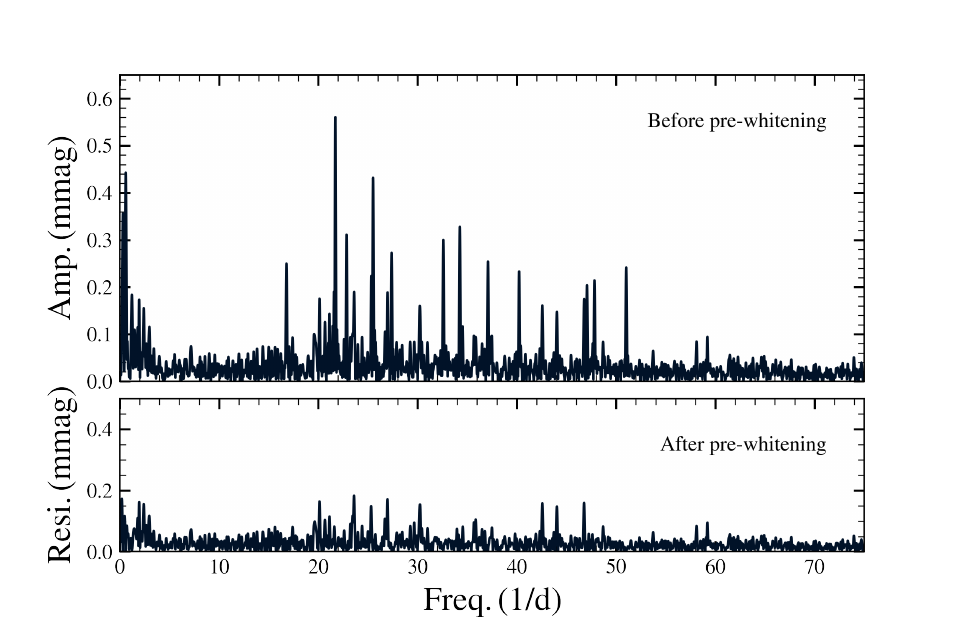}
  \caption{TIC~165459595}
  \label{fig:TIC~16545_period}
\end{subfigure}
\hfill
\begin{subfigure}{0.49\textwidth}
  \centering
  \includegraphics[width=\linewidth]{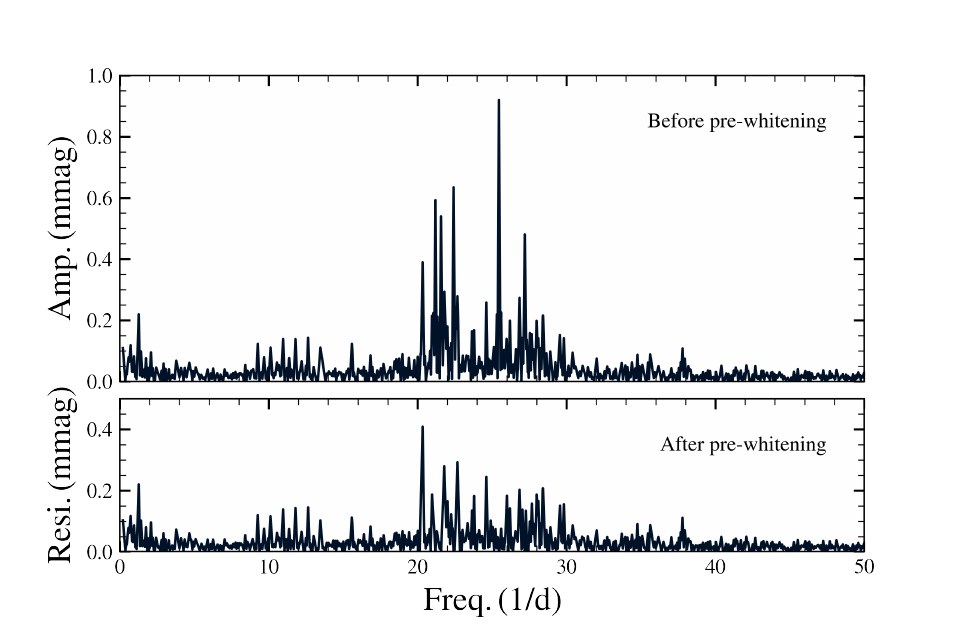}
  \caption{TIC$~$308953703}
  \label{fig:TIC~3089_period}
\end{subfigure}

\caption{Periodograms for residuals obtained after subtracting \textsc{jktebop} models from the corresponding LCs.}
\end{figure*}

The target was selected based on the stellar masses of the two stars. Although the LC did not show clear evidence of pulsations, initial checks did hint towards a couple of very low-amplitude pulsation signals. The binary subtracted residuals, however, show two very faint peaks (S/N $\sim$ 3; see Fig. \ref{fig:TIC~18978_period}) that are extremely close to the noise floor. These peaks fall below our set S/N limit. We therefore mark this as a false positive. 

This target has 2-min cadence data only from a single \textsc{tess} sector so far. More photometry will help in boosting the S/N, which is necessary to test the nature of these low-amplitude frequency peaks.


\subsection{TIC~386622782}
The large number of independent and combination frequencies, mentioned in Table~\ref{tab:HR2214_freqs}, result in the complex pulsations visible in the LC. Considering the large flux contribution from the secondary, it is most probable that it is the pulsating component. However, it cannot be ruled out that the other star may also be pulsating, although the relative contribution of those pulsations to the visible frequency spectra would be minimal.


\subsection{TIC~165459595}
The frequencies for this target fall in the range of $\delta$ Scuti-type oscillations. Stars with an initial mass greater than 1.5 times $M\mathrm{_{sun}}$ can cross the instability strip before entering the MS \citep{2009AIPC.1170..403H}. Such oscillations can therefore be explained despite of the young age of the system.
\par
The $T\mathrm{_{eff}}$ and $\log(g)$ for $\delta$ Scuti-type stars typically have values between 6300-8500 K and 3.2-4.3, respectively, corresponding to the A-F type variables. However, for this target the $T\mathrm{_{eff}}$ estimates from the isochrone fits and the $\log(g)$ from the LC modelling would suggest a B-type classification. One such candidate with higher $T\mathrm{_{eff}}$, AL~Scl, was recently found by \citet{2023MNRAS.524..619K}. Better constraints on the atmospheric parameters are necessary to explain the pulsations observed in this case.


\subsection{TIC$~$308953703}
The extracted independent frequencies, $f\mathrm{_1}$ and $f\mathrm{_3}$, and their amplitudes are close to the values obtained by \citet{2021MNRAS.508..529F}. Since they have already performed a detailed asteroseismic study for this target, we have not attempted any further analysis.

\begin{figure}[htbp]
         \centering
         \includegraphics[width=0.495\textwidth]{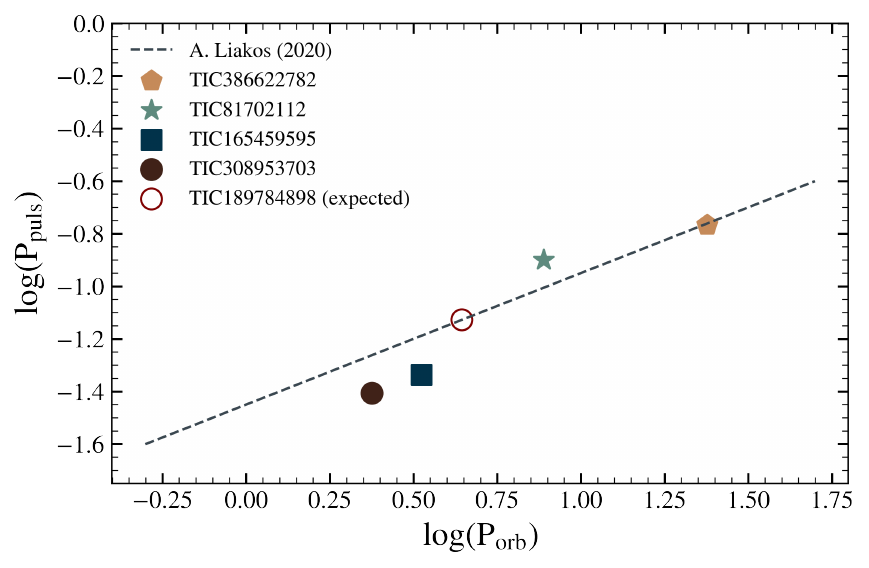}
         \caption{Location of the dominant pulsating frequency for the studied systems relative to the $P_\mathrm{{orb}}$-$P_\mathrm{{pul}}$ empirical relation (dotted line) derived by \citet{2020A&A...642A..91L}, for such systems with $P_\mathrm{{orb}}$ < 13 d.} 
\label{fig:PP_relation}
\end{figure}

Using the $P_\mathrm{{orb}}$ values from LC modelling and the period corresponding to the highest-amplitude frequency, we located our targets on the log$P_\mathrm{{orb}}$-log$P_\mathrm{{pul}}$ diagram (see Fig. \ref{fig:PP_relation}). We overplotted the $P_\mathrm{{orb}}$-$P_\mathrm{{pul}}$ empirical relation obtained by \citet{2020A&A...642A..91L}, and used it check the expected $P_\mathrm{{pul}}$ for TIC~189784898.

\section{Conclusions}

We have presented a comprehensive analysis for four candidate DEBs with at least one component exhibiting $\delta$ Scuti-type pulsations. We have confirmed the presence of pulsations in three systems, and obtained a set of precise parameters for both stars. We have also presented updated parameters for TIC$~$308953703 based on the LC modelling of the short-cadence TESS photometry.

The variability of the pulsation frequency in the case of TIC~81702112 makes this target an extremely interesting case study. However, detailed modelling to explain the mechanism responsible for this is beyond the scope of this paper. It is necessary to use a combination of techniques, such as multi-band photometry for mode identification, detailed binary evolution models, and asteroseismic modelling, to obtain a better understanding of this unique target.

We are able to put good constraints on the physical properties and evolutionary stage of all the targets in the sample except for TIC~165459595. The main limitation in this case is the unavailability of high-quality spectra. However, the obtained masses, radii, and flux ratio were enough to estimate the stellar ages in the range of a few million years. Our understanding of $\delta$ Scuti-type pulsations in pre-MS stars is still in the nascent stages. Observational studies of such targets are crucial to aid the ongoing theoretical studies \citep{2021A&A...654A..36S, 2023MNRAS.526.3779M} of such stars.

Building upon the detailed study by \citet{2021MNRAS.508..529F}, we have been able to update the physical parameters for TIC$~$308953703 and prove that the evolutionary stages of both stars can in fact be explained within a co-evolution scenario. This study attempts to further enrich the sample of well-studied $\delta$ Scuti-type pulsators in EBs. Such a sample is crucial to understand the extent of the effect of binarity on stellar pulsations. With high-quality space photometry, the detection of such systems is rising rapidly. This presents a great opportunity to use these unique and powerful tools to study the structure and evolution of such stars in great detail.

\section*{Acknowledgements}

T.P. would like to thank A. Miszuda for the fruitful discussions and suggestions. T.P. acknowledges the support provided by the Polish National Science Centre (NCN) through grant no. 2021/43/B/ST9/02972 and 2017/27/B/ST9/02727. K.G.H. acknowledges the support provided by the NCN with the grant 2023/49/B/ST9/01671. A.M. acknowledges the support provided by the NCN with the grant 2021/41/N/ST9/02746. This research made use of Lightkurve, a Python package for Kepler and TESS data analysis \citep{2018ascl.soft12013L}. This research uses data collected by the TESS mission, which are publicly available from the Mikulski Archive for Space Telescopes (MAST) at the Space Telescope Science Institute (STScI). Funding for the TESS mission is provided by NASA’s Science Mission directorate.

%
%


\bibliographystyle{aa}
\bibliography{pawar}

\begin{appendix}
\label{appendix}

\section{\textsc{jktebop} MC runs}

\begin{figure*}[h!]
         \centering
         \includegraphics[width=0.95\textwidth]{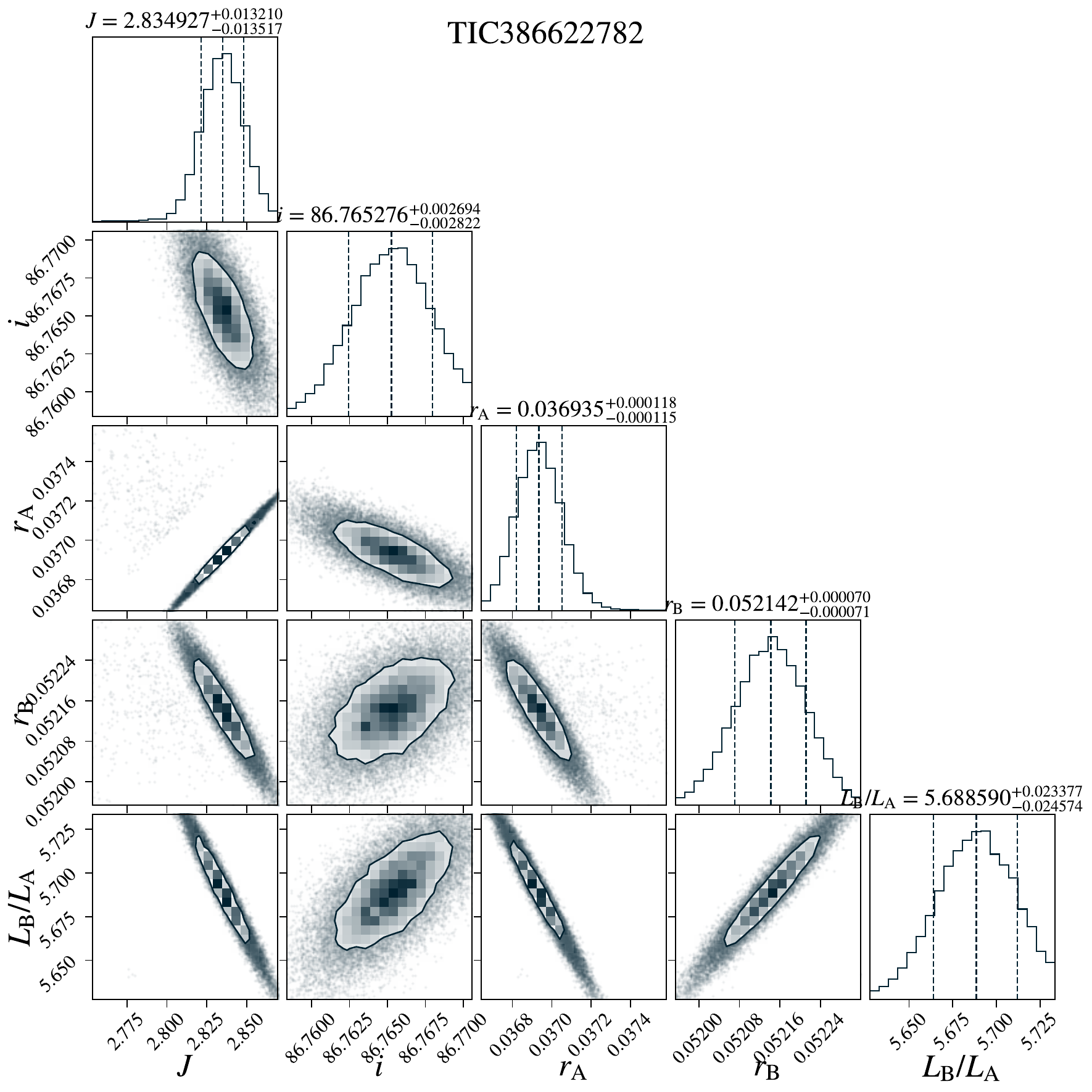}
         \caption{Corner plot for the distribution of parameter values from the MC runs within \textsc{jktebop}. For $i$ we use 95\% of the sampling to avoid values where the secondary eclipse is left undetected.} 
         \label{fig:HR2214_mc_model}
\end{figure*}

\begin{figure*}[h!]
         \centering
         \vspace{1cm}
         \includegraphics[width=0.495\textwidth]{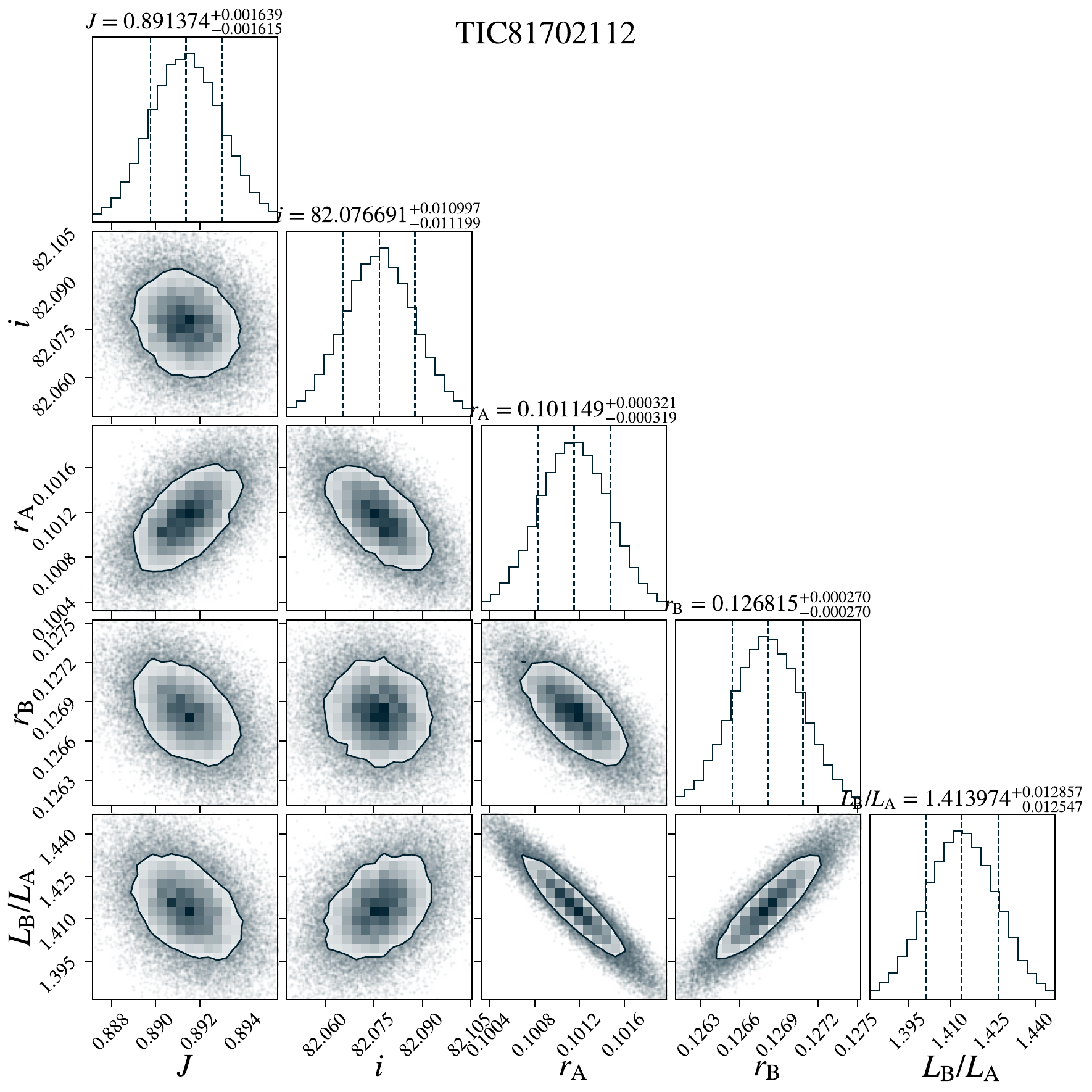}
         \includegraphics[width=0.495\textwidth]{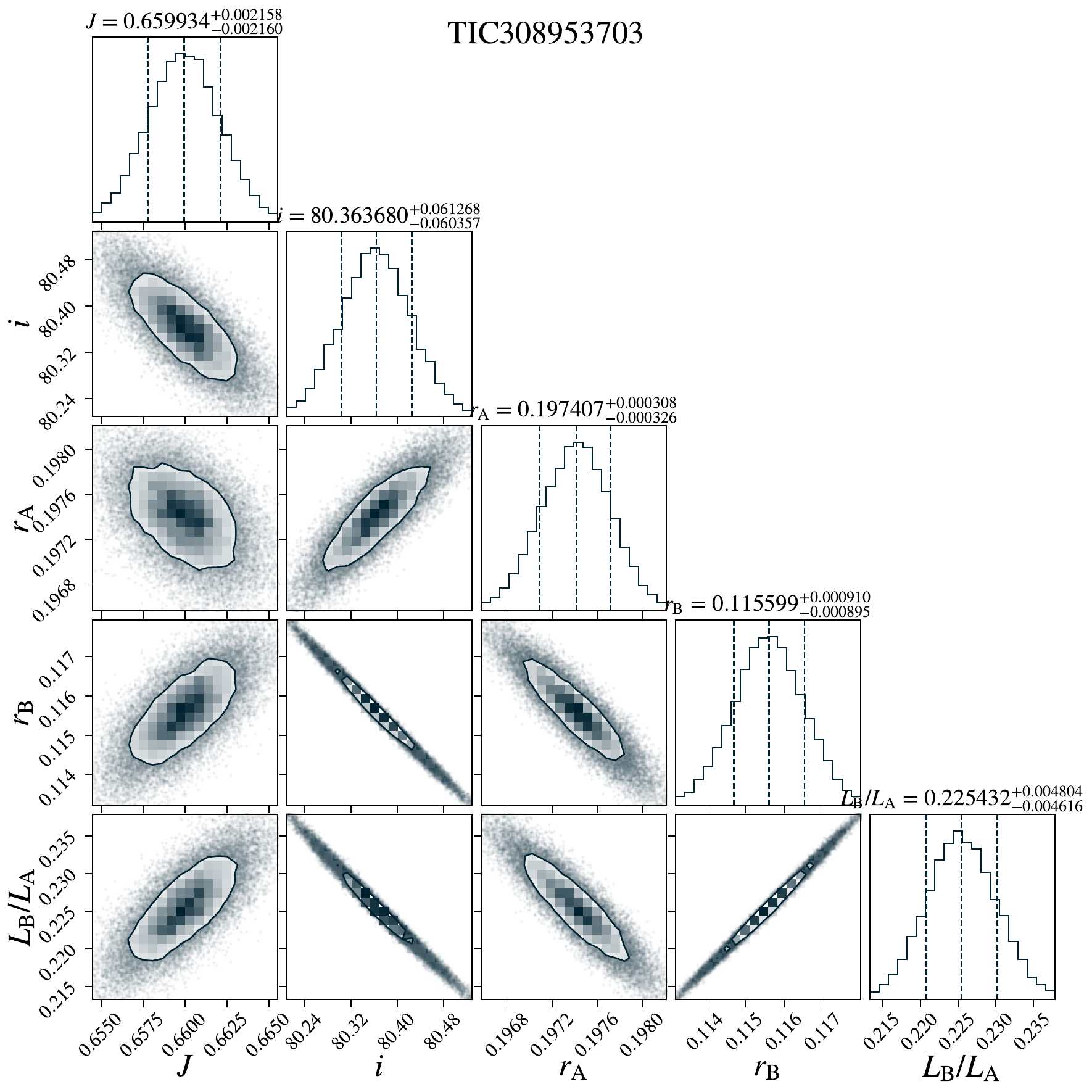}\\
         \vspace{1cm}
         \includegraphics[width=0.495\textwidth]{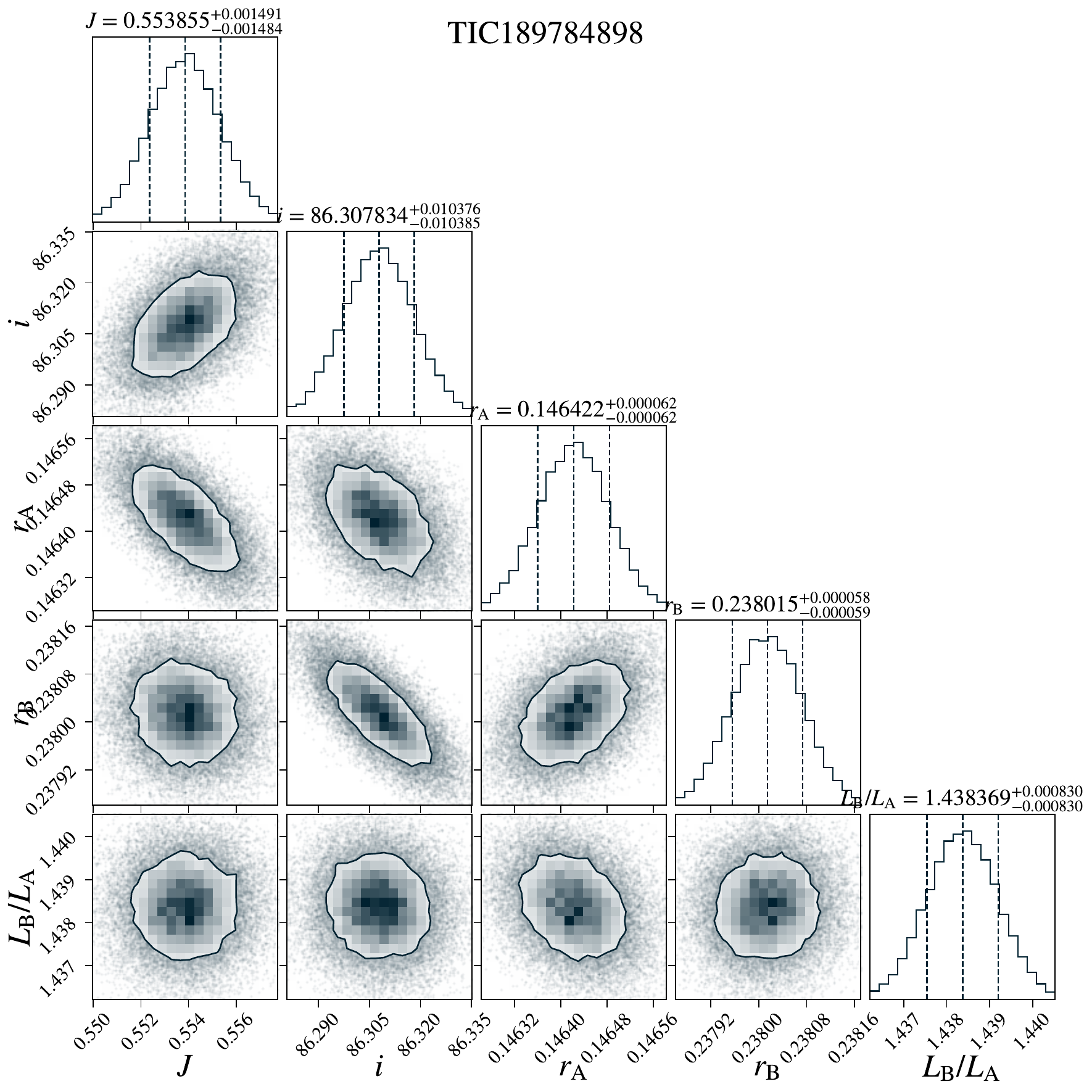}
         \includegraphics[width=0.495\textwidth]{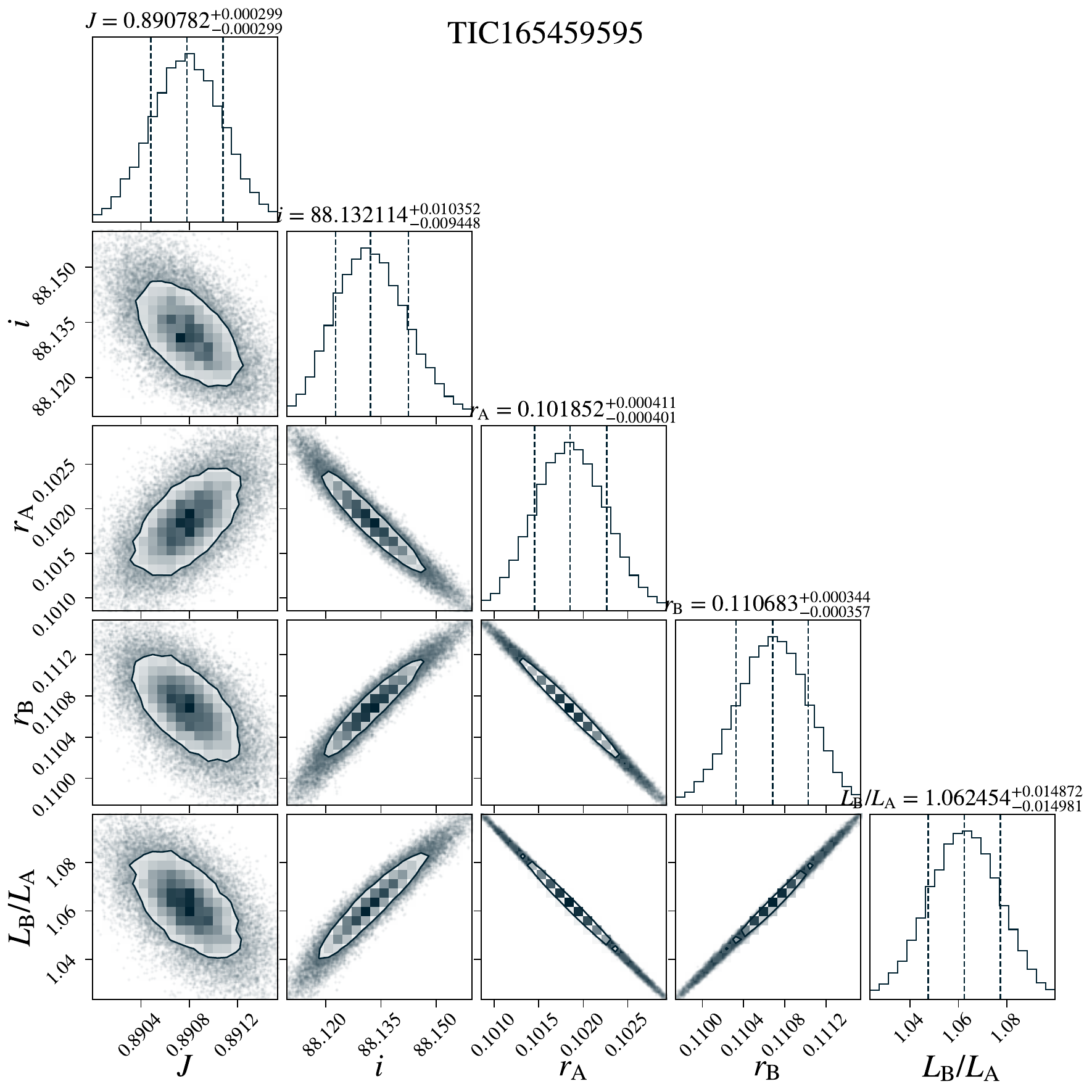}
         \caption{Corner plots for the other targets corresponding to their best-fit \textsc{jktebop} models.}
         \label{fig:Corner_plots}
\end{figure*}

\section{Additional Table}

\begin{sidewaystable*}[h!] 
    \centering
\small
    \caption{Parameters of the four EB systems obtained using a combination of RV-LC modelling, spectral analysis and isochrone fitting.}
    \label{tab:Final_parameters}
    \begin{tabular}{lccccc}
    
       \textbf{Parameters} & \textbf{TIC~81702112} & \textbf{TIC~189784898} & \textbf{TIC~386622782} & \textbf{TIC~165459595} & \textbf{\textbf{TIC$~$308953703}} \\
        \hline
        $t\mathrm{_o}$ (BJD-2450000)         & 10019.652801 $\pm$ 0.000128 & 9336.375085 $\pm$ 0.000018 & 9478.4924936103 (fixed)  & 9312.145783 $\pm$  0.000012 & 9892.05091 $\pm$ 0.000043 \\
        $P_\mathrm{orb}$ (days)       & 7.735799 $\pm$ 0.000028     & 4.409041 $\pm$ 0.000010    & 23.809526 $\pm$ 0.000001 & 3.336960 $\pm$ 0.000011     & 2.372223 $\pm$ 0.000022 \\
        $r\mathrm{_A}+r\mathrm{_B}$   & 0.22797 $\pm$ 0.00024       & 0.37853 $\pm$ 0.00030      & 0.08908 $\pm$ 00006      & 0.21255 $\pm$ 0.00005       & 0.31301 $\pm$ 0.00064 \\
        $r\mathrm{_B}$/$r\mathrm{_A}$ & 1.25375 $\pm$ 0.00616       & 1.68373 $\pm$ 0.00148      & 1.41185 $\pm$ 0.00625    & 1.08668 $\pm$ 0.00571       &  0.53561 $\pm$ 0.00545 \\
        $J$                           & 0.8914 $\pm$ 0.0016         & 0.5173 $\pm$ 0.0011        & 2.8349 $\pm$ 0.0135      & 0.8907 $\pm$ 0.0003         & 0.6599 $\pm$ 0.0022 \\
        $i$ ($\mathrm{^o}$)           & 82.077 $\pm$ 0.011          & 86.308 $\pm$ 0.010         & 86.765 $\pm$ 0.003       & 88.132 $\pm$ 0.009          & 80.360 $\pm$ 0.068 \\
        $q$                           & 1.05669                     & 1.0345 $\pm$ 0.0051        & 1.1899 $\pm$ 0.0029      & 1.0857 $\pm$ 0.0292         & 0.7795  \\
        $e$                           & 0.20715 (fixed)             & 0 (fixed)                  & 0.5424 (fixed)           & 0 (fixed)                   & 0.02697 $\pm$ 0.0009 \\
        $\omega$ ($\mathrm{^o}$) & 85.26 $\pm$ 0.06 (\textsc{v2fit}) & 0 (fixed)                  & 91.35 $\pm$ 0.02         & 0 (fixed)                   & 90.0 (fixed) \\
        $K_1$ [km\,s$^{-1}$]          & 85.52 $\pm$ 0.11            & 97.45 $\pm$ 0.09           & 75.03 $\pm$ 0.15         & 130.41 $\pm$ 2.5            & 98.20 $\pm$ 0.7 \\
        $K_2$ [km\,s$^{-1}$]    & 80.01 $\pm$ 0.10 (\textsc{v2fit}) & 93.70 $\pm$ 0.42           & 63.06 $\pm$ 0.07         & 120.11 $\pm$ 2.2            & 122.14 $\pm$ 1.0 \\
        $\gamma$ [km\,s$^{-1}$] & 10.93 $\pm$ 0.08   &  6.53 $\pm$ 0.12 & 36.15 $\pm$ 0.04   &  9.311 $\pm$ 0.1 &  -14.65 $\pm$ 1.5 \\
        $L\mathrm{_B}$/$L\mathrm{_A}$ & 1.41403 $\pm$ 0.0128        & 1.4384 $\pm$ 0.0004        & 5.6886 $\pm$ 0.0241      & 1.0624 $\pm$ 0.0150         & 0.2255 $\pm$ 0.005 \\
        $r\mathrm{_A}$                & 0.10115 $\pm$ 0.00032       & 0.14642 $\pm$ 0.00004      & 0.03694 $\pm$ 0.00012    & 0.10185 $\pm$ 0.00040       & 0.19741 $\pm$ 0.00030 \\
        $r\mathrm{_B}$                & 0.12682 $\pm$ 0.00027       & 0.23802 $\pm$ 0.00005      & 0.05214 $\pm$ 0.00007    & 0.11068 $\pm$ 0.00035       & 0.11560 $\pm$ 0.00091 \\
        $a$ [R$_\odot$]               & 25.543 $\pm$ 0.022     & 16.708 $\pm$ 0.038      & 54.687 $\pm$ 0.063   & 16.51 $\pm$ 0.2           & 10.333 $\pm$ 0.034 \\
        $M_\mathrm{A}$ [M$_\odot$]    & 1.808 $\pm$ 0.005           & 1.578 $\pm$ 0.005          & 1.768 $\pm$ 0.005        & 2.611 $\pm$ 0.11            & 1.478 $\pm$ 0.027 \\
        $M_\mathrm{B}$ [M$_\odot$]    & 1.933 $\pm$ 0.005           & 1.642 $\pm$ 0.014          & 2.103 $\pm$ 0.008        & 2.834 $\pm$ 0.12            & 1.152 $\pm$ 0.018 \\
        $R_\mathrm{A}$ [R$_\odot$]    & 2.585 $\pm$ 0.009           & 2.357 $\pm$ 0.006          & 2.020 $\pm$ 0.007        & 1.684 $\pm$ 0.024           & 2.039 $\pm$ 0.012 \\
        $R_\mathrm{B}$ [R$_\odot$]    & 3.239 $\pm$ 0.008           & 3.968 $\pm$ 0.009          & 2.852 $\pm$ 0.005        & 1.830 $\pm$ 0.025           & 1.195 $\pm$ 0.011 \\
        $\log(g)_\mathrm{A}$             & 3.871 $\pm$ 0.003           & 3.892 $\pm$ 0.002          & 4.075 $\pm$ 0.003        & 4.366 $\pm$ 0.008           & 3.988 $\pm$ 0.001 \\
        $\log(g)_\mathrm{B}$             & 3.703 $\pm$ 0.002           & 3.456 $\pm$ 0.001          & 3.851 $\pm$ 0.002        & 4.366 $\pm$ 0.008           & 4.435 $\pm$ 0.016 \\
        Distance [pc]    (J-band)       & 302 $\pm$ 7 & 575 $\pm$ 23 & 103 $\pm$ 3 & 421 $\pm$ 11 & 292 $\pm$ 7 \\
        Age (Gyr)            & 1.125 $\pm$ 0.025  & 1.9 $\pm$ 0.05  & 0.8 $\pm$ 0.05 & 0.1  $\pm$ 0.2 & 1.75 $\pm$ 0.25 \\
        \hline
    \end{tabular}

\end{sidewaystable*}
\section{Pulsation Frequencies}
\begin{table*}[h!]
\centering
\small
\caption{Independent and combination frequencies for TIC~81702112}
\label{tab:VCir_freqs}

\begin{tabular}{lcccc}

Label & Frequency & Amplitude & Phase & Comment \\
\hline
F1 & 7.9453259  $\pm$ 0.0000357 & 0.0079663 $\pm$ 0.0000274 & 0.8030060 $\pm$ 0.0005363 & - \\
F2 & 7.6890551  $\pm$ 0.0000828 & 0.0034384 $\pm$ 0.0000284 & 0.5265828 $\pm$ 0.0013870 & - \\
F3 & 7.8428176  $\pm$ 0.0002846 & 0.0010009 $\pm$ 0.0000275 & 0.4040479 $\pm$ 0.0042617 & - \\
F4 & 7.7058292  $\pm$ 0.0003301 & 0.0008631 $\pm$ 0.0000280 & 0.2415520 $\pm$ 0.0035554 & $\sim$f2 \\
F5 & 0.0354120  $\pm$ 0.0004158 & 0.0006855 $\pm$ 0.0000288 & 0.8746073 $\pm$ 0.0078626 & $\sim$ f4-f2 \\
F6 & 7.1839687  $\pm$ 0.0003704 & 0.0007691 $\pm$ 0.0000263 & 0.5778487 $\pm$ 0.0056428 & $\sim$ 3f2-2f1 \\
F7 & 0.4659469  $\pm$ 0.0004078 & 0.0006980 $\pm$ 0.0000279 & 0.2227401 $\pm$ 0.0062876 & $\sim$ f2-f5-f6 \\
F8 & 0.0130465  $\pm$ 0.0006036 & 0.0004728 $\pm$ 0.0000565 & 0.9302906 $\pm$ 0.0201806 & $\sim$ f5 \\
F9 & 15.6334491 $\pm$ 0.0006289 & 0.0004529 $\pm$ 0.0000283 & 0.9889682 $\pm$ 0.0097948 & $\sim$ f1+f2 \\
F10 & 0.4883123 $\pm$ 0.0006000 & 0.0004756 $\pm$ 0.0000290 & 0.8754320 $\pm$ 0.0092762 & $\sim$ f7 \\
F11 & 0.2982060 $\pm$ 0.0006110 & 0.0004658 $\pm$ 0.0000270 & 0.8357743 $\pm$ 0.0091506 & $\sim$ f1+f5-f2 \\
F12 & 15.8906517 $\pm$ 0.0006554 & 0.0004347 $\pm$ 0.0000275 & 0.9692720 $\pm$ 0.0098979 & $\sim$ 2f1 \\
F13 & 15.2429856 $\pm$ 0.0007236 & 0.0003938 $\pm$ 0.0000300 & 0.5279298 $\pm$ 0.0137924 & $\sim$ f2+f3-f11 \\
F14 & 0.1798555 $\pm$ 0.0007852 & 0.0003623 $\pm$ 0.0000274 & 0.6694213 $\pm$ 0.0123499 & $\sim$ f3-f2 \\
F15 & 0.8126113 $\pm$ 0.0008032 & 0.0003548 $\pm$ 0.0000275 & 0.6350222 $\pm$ 0.0133010 & $\sim$ f10+f11 \\
\hline
\end{tabular}

\label{tab:A111134_freqs}
\end{table*}


\begin{table*}[h!]
\centering
\small
\caption{Independent and combination frequencies for TIC~386622782}
\label{tab:HR2214_freqs}
\begin{tabular}{lcccc}

Label & Frequency & Amplitude & Phase & Comment \\
\hline
F1 & 5.8561934  $\pm$     0.0000281& 0.0006582 $\pm$ 0.0000020 & 0.5424589 $\pm$ 0.0005057 & - \\
F2 & 9.1214054  $\pm$     0.0000424& 0.0004263 $\pm$ 0.0000026 & 0.7381891 $\pm$ 0.0023635 & - \\
F3 & 8.4066294  $\pm$     0.0000707& 0.0002636 $\pm$ 0.0000021 & 0.5658486 $\pm$ 0.0012279 & - \\
F4 & 6.8233972  $\pm$     0.0000840& 0.0002194 $\pm$ 0.0000020 & 0.3864127 $\pm$ 0.0014482 & - \\
F5 & 15.60082073 $\pm$    0.0000937& 0.0001983 $\pm$ 0.0000020 & 0.2679418 $\pm$ 0.0016001 & - \\
F6 & 7.3812233  $\pm$     0.0001003& 0.0001879 $\pm$ 0.0000021 & 0.4853742 $\pm$ 0.0016767 & - \\
F7 & 8.9252180  $\pm$     0.0001049& 0.0001799 $\pm$ 0.0000021 & 0.5812259 $\pm$ 0.0018364 & $\sim$ 2f6-f1 \\
F8 & 8.0181784  $\pm$     0.0001055& 0.0001785 $\pm$ 0.0000021 & 0.2089677 $\pm$ 0.0018900 & $\sim$ f4+2f3-f5 \\
F9 & 18.39583706 $\pm$    0.0001225& 0.0000226 $\pm$ 0.0000140 & 0.3182499 $\pm$ 0.0550578 & $\sim$ f5+2f1-f7 \\
F10 & 8.44325107 $\pm$    0.0001292& 0.0001437 $\pm$ 0.0000021 & 0.1358174 $\pm$ 0.0021979 & $\sim$ 3f1-f2 \\
F11 & 2.72896658 $\pm$    0.0001413& 0.0001058 $\pm$ 0.0001187 & 0.6557422 $\pm$ 0.1950036 & $\sim$ 3f7-3f8 \\
F12 & 7.54275098 $\pm$    0.0001529& 0.0001178 $\pm$ 0.0000020 & 0.7056628 $\pm$ 0.0027418 & $\sim$ f2+f4-f3 \\
F13 & 6.61086095 $\pm$    0.0002101& 0.0000895 $\pm$ 0.0000020 & 0.6410822 $\pm$ 0.0037328 & $\sim$ f4+f7-f2 \\
F14 & 4.69149430 $\pm$    0.0003727& 0.0000550 $\pm$ 0.0000021 & 0.7998891 $\pm$ 0.0058628 & $\sim$ f11+f5-2f4 \\
F15 & 3.64646955 $\pm$    0.0003972& 0.0000445 $\pm$ 0.0000020 & 0.8010967 $\pm$ 0.0072520 & $\sim$ f2-2f11 \\
F16 & 9.13056084 $\pm$    0.0003739& 0.0000564 $\pm$ 0.0000026 & 0.8876553 $\pm$ 0.0013271 & $\sim$ f2 \\
F17 & 8.54330668 $\pm$    0.0004023& 0.0000478 $\pm$ 0.0000049 & 0.4906313 $\pm$ 0.0457402 & $\sim$ f7+f8-f3 \\
F18 & 8.21840914 $\pm$    0.0004964& 0.0000374 $\pm$ 0.0000024 & 0.4329490 $\pm$ 0.0108088 & $\sim$ f5-f6 \\
F19 & 7.48384239 $\pm$    0.0005025& 0.0000369 $\pm$ 0.0000026 & 0.9078710 $\pm$ 0.0109404 & $\sim$ f1+f10-f4 \\

\hline
\end{tabular}

\end{table*}


\begin{table*}[h!]
\centering
\small
\caption{Independent and combination frequencies for TIC~165459595}
\label{tab:TIC~165459595_freqs}
\begin{tabular}{lcccc}

Label & Frequency & Amplitude & Phase & Comment \\
\hline
F1 & 21.7250931  $\pm$ 0.0022925  & 0.0005542 $\pm$ 0.0000184 & 0.3455254 $\pm$ 0.0051434 & - \\
F2 & 0.5977860   $\pm$ 0.0024522  & 0.0005085 $\pm$ 0.0000184 & 0.8216044 $\pm$ 0.0055903 & $\sim$ 2$f_\mathrm{{orb}}$ \\
F3 & 25.5147602  $\pm$ 0.0030086  & 0.0004299 $\pm$ 0.0000180 & 0.1016392 $\pm$ 0.0068261 & - \\
F4 & 34.2601479  $\pm$ 0.0039663  & 0.0003257 $\pm$ 0.0000188 & 0.2270885 $\pm$ 0.0091310 & - \\
F5 & 0.2988930   $\pm$ 0.0040300  & 0.0003026 $\pm$ 0.0000178 & 0.4087414 $\pm$ 0.0096954 & $\sim$ f3+2f2-2f4 \\
F6 & 32.5830261  $\pm$ 0.0042263  & 0.0003014 $\pm$ 0.0000191 & 0.9674358 $\pm$ 0.0101209 & $\sim$ f4-3f2 \\
F7 & 22.8431743  $\pm$ 0.0044417  & 0.0002982 $\pm$ 0.0000188 & 0.8440174 $\pm$ 0.0101409 & $\sim$ f1+2f2 \\
F8 & 27.3800739  $\pm$ 0.0046920  & 0.0002745 $\pm$ 0.0000189 & 0.6419112 $\pm$ 0.0107901 & $\sim$ f3+3f2 \\
F9 & 37.0940964  $\pm$ 0.0051350  & 0.0002545 $\pm$ 0.0000199 & 0.4846070 $\pm$ 0.0150174 & $\sim$ f6+f8-f7 \\
F10 & 40.2380079 $\pm$ 0.0055421  & 0.0002336 $\pm$ 0.0000183 & 0.6635547 $\pm$ 0.0126312 & $\sim$ f1+2f3-f6 \\
\hline
\end{tabular}

\end{table*}


\begin{table*}[h!]
\centering
\small
\caption{Independent and combination frequencies for TIC$~$308953703}
\label{tab:TIC~308953703_freqs}
\begin{tabular}{lcccc}

Label & Frequency & Amplitude & Phase & Comment \\
\hline
F1&	25.4700692  $\pm$     0.0011920  &    0.000911  $\pm$ 0.0000192 & 0.8323153 $\pm$ 0.0035427 & -  \\
F2&	22.4233029  $\pm$     0.0017536  &    0.000627  $\pm$ 0.0000199 & 0.2705903 $\pm$ 0.0052324 & -  \\
F3&	21.2031232  $\pm$     0.0019406  &    0.000567  $\pm$ 0.0000194 & 0.9900451 $\pm$ 0.0057373 & -  \\
F4&	21.5831792  $\pm$     0.0021572  &    0.000504  $\pm$ 0.0000203 & 0.9297193 $\pm$ 0.0065013 & -  \\
F5&	27.2003241  $\pm$     0.0022049  &    0.000494  $\pm$ 0.0000198 & 0.7133185 $\pm$ 0.0063090 & $\sim$ f3+2f1-2f2 \\
F6&	0.4550670   $\pm$     0.0027142  &    0.000400  $\pm$ 0.0000201 & 0.0560345 $\pm$ 0.0079532 & $\sim$ $f_\mathrm{{orb}}$  \\
F7&	20.3379957  $\pm$     0.0026604  &    0.000407  $\pm$ 0.0000195 & 0.5658421 $\pm$ 0.0077038 & $\sim$ f3+f4-f2 \\
\hline
\end{tabular}

\end{table*}


\end{appendix}
\end{document}